%% file: main.tex
\documentclass[aps,prd,twoside,singlecolumn,superscriptaddress,floatfix,nofootinbib]{revtex4}
\usepackage[utf8]{inputenc}
\usepackage[english]{babel}
\usepackage{natbib}

\usepackage{physics}
\usepackage{graphicx}
\usepackage{esint}
\usepackage{tabularx} % extra features for tabular environment
\usepackage{amsmath}  % improve math presentation
\usepackage{graphicx} % takes care of graphic including machinery
\usepackage[margin=.75in,letterpaper]{geometry} % decreases margins
\usepackage[final]{hyperref} % adds hyper links inside the generated pdf file
\hypersetup{
	colorlinks=true,       % false: boxed links; true: colored links
	linkcolor=blue,        % color of internal links
	citecolor=blue,        % color of links to bibliography
	filecolor=magenta,     % color of file links
	urlcolor=blue         
}

\begin{document}
\title{Comparison of Halo Model and Simulation Predictions for Projected-Field Kinematic Sunyaev-Zel'dovich Cross-Correlations}

\author{Michael Jacob Rodriguez}
\affiliation{Department of Physics, Columbia University, New York, NY, USA 10027}

\author{Aleksandra Kusiak}
\affiliation{Kavli Institute for Cosmology, Cambridge, Madingley Road, Cambridge CB3 0HA, UK}
\affiliation{Institute of Astronomy, University of Cambridge, Madingley Road, Cambridge, CB3 0HA, UK}
\affiliation{Department of Physics, Columbia University, New York, NY, USA 10027}

\author{Shivam Pandey}
\affiliation{William H.~Miller III Department of Physics \& Astronomy, Johns Hopkins University, Baltimore, MD 21218, USA}
\affiliation{Department of Physics, Columbia University, New York, NY, USA 10027}

\author{J.~Colin Hill}
\affiliation{Department of Physics, Columbia University, New York, NY, USA 10027}

\begin{abstract}
The kinematic Sunyaev-Zel’dovich (kSZ) effect in the cosmic microwave background (CMB) is a powerful probe of gas physics and large-scale structure (LSS) in our universe. Here we consider the ``projected-field'' kSZ estimator, which involves cross-correlating a foreground-cleaned, filtered, squared CMB temperature map with an LSS tracer, and requires no individual tracer redshifts.  We compare \verb|class_sz| halo model calculations of projected-field kSZ cross-correlations with measurements of these signals from the Websky numerical simulations.  We extract dark matter halo catalogs from Websky and cross-correlate halo number density maps with various CMB secondary signals.  We first validate our halo model by comparing its predictions for thermal SZ (tSZ) and patchy screening ($\tau$) cross-correlations to measurements of these signals from Websky.  We consider three different halo redshift ranges in our comparisons.  We also construct our own kSZ, tSZ, and $\tau$ maps to validate the form of the relevant profiles.  Following the tSZ and $\tau$ validation, we compare projected-field kSZ calculations between the halo model and the simulations.  We use filters constructed for \textit{Planck} and the Simons Observatory (SO) to assess the accuracy of the halo-model kSZ predictions for experiments of differing sensitivity.  %We compare our results with previous projected-field works to see the significance of other terms in the kSZ five-point cross-correlation and forecast our results with a larger halo sample size than originally present in Websky. 
Overall, we find reasonable agreement, particularly at \emph{Planck} sensitivity.  However, we find an $\approx$ 50$\%$ difference between our halo model and the simulation measurements for SO, which significantly exceeds the predicted error bars.  We note that our halo model includes only the dominant expected term in the projected-field kSZ signal; the magnitude of the difference between our model and the simulations is consistent with previous predictions for terms arising from other contractions in the theory calculation.  These terms will need to be included to obtain unbiased inference from upcoming projected-field kSZ measurements.  
\end{abstract}

\maketitle

\section{Introduction}
\label{sec:intro}

The kinematic Sunyaev-Zel’dovich (kSZ) effect is the Doppler shift of cosmic microwave background (CMB) photons due to Compton scattering off of ionized electrons (e.g., in and around galaxies) moving with non-zero bulk velocity along the line of sight (LOS), which is sensitive to the momentum of the electrons~\cite{1980ARA&A..18..537S}.  The thermal SZ (tSZ) effect is the inverse-Compton-scattering-induced boosting of CMB photons, which is sensitive to the electron pressure~\cite{1969Natur.223..721S}. These effects probe large-scale structure (LSS) and are important in studying gas physics involved in galaxy formation and evolution~\cite{2019BAAS...51c.297B}. However, there are several challenges involved in measuring the kSZ signal. For one, the CMB blackbody spectrum is preserved in the kSZ field, meaning a multi-frequency analysis cannot separate the kSZ signal from the primary CMB, unlike the tSZ effect, which is a frequency-dependent phenomenon. The kSZ effect is also only detectable on very small angular scales, meaning high-resolution data is necessary to detect it. A few methods seek to work around these difficulties to detect the kSZ signal. These include methods like velocity-weighted stacking {\cite{hadzhiyska2025evidencelargebaryonicfeedback, Guachalla_2025}, which stacks cut-outs of CMB maps centered at galaxy locations weighted by the LOS velocity field reconstructed from spectroscopic galaxy surveys, and the mean pairwise momentum method \cite{ferreira1999measuringomegagalaxystreaming}, which leverages the fact that the average pairwise momenta of galaxies are negative when $\sim25-50$ Mpc apart, resulting from gravity \cite{hadzhiyska2025probingcosmicvelocitiespairwise, gong2025detectionpairwisekinematicsunyaevzeldovich}. Another kSZ estimator that has seen significant recent development is the velocity reconstruction technique, which measures the large-scale velocity field using small-scale kSZ-galaxy cross-correlations, e.g., \cite{mccarthy2024atacamacosmologytelescopelargescale, laguë2024constraintslocalprimordialnongaussianity, hotinli2025velocityreconstructionkszmeasuring, mccarthy2025atacamacosmologytelescopecrosscorrelation}. All three of these methods probe the $\langle Tgg \rangle$ bispectrum, where $T$ is the kSZ temperature field and $g$ denotes an LSS tracer (e.g., a galaxy)~\cite{2018arXiv181013423S}. In this work, the method that will be of focus is the projected-field method, which involves cross-correlating the square of the kSZ field with an LSS tracer. This method involves first performing extensive multi-frequency foreground cleaning to remove non-blackbody CMB signals, followed by filtering the CMB map with an angular-scale-dependent filter in order to upweight the kSZ signal relative to the primary CMB.  This estimator probes the $\langle TTg\rangle$ bispectrum.  Importantly, the $\langle Tgg \rangle$-based methods require redshift measurements of each LSS tracer (e.g., the mean pairwise momentum procedure requires the redshift of each tracer to precisely obtain the distances between galaxies), although, as shown in several works \cite[e.g.,][]{2016MNRAS.461.3172S,hadzhiyska2025evidencelargebaryonicfeedback}, the per-object redshift estimates do not have to spectroscopic. In contrast, the projected-field approach considers only \emph{projected} 2D maps, and thus does not require per-object redshift information. In fact, it requires only the overall redshift distribution of the tracer sample, making it applicable to catalogs that do not provide per-object redshifts at all, as well as to other LSS tracers, such as CMB or galaxy lensing \cite{Dore2004,Bolliet_2023}, unlike other kSZ estimators.

The projected-field kSZ estimator was first proposed in \cite{Dore2004, DeDeo}, and applied to data for the first time in \cite{Hill2016,Ferraro2016} using the \emph{Planck} and \emph{Wide-Field Infrared Surveyor} (\emph{WISE}) data, resulting in a 4$\sigma$ measurement.  The second measurement was performed using the same \emph{Planck} maps and the \emph{unWISE} galaxy catalog \cite{krolewski_2020} in \cite{Kusiak_2021}. Both measurements used an ``effective approach'' to model the signal, based on a combination of perturbation theory approximations and fitting functions from numerical simulations.  In this approach, one first notes that the dominant term in the projected-field $\langle TTg\rangle$ bispectrum can be written as the product of the large-scale velocity dispersion and the non-linear matter bispectrum, $B_{\delta_g {p_{ \boldsymbol{\hat{n}}} p_{\boldsymbol{\hat{n}}}}} = \frac{1}{3}v_\mathrm{rms}^2 B_\mathrm{m}^\mathrm{NL}$.  A numerical fitting function is then used to model the non-linear matter bispectrum $B_\mathrm{m}^\mathrm{NL}$, with the electrons assumed to trace the dark matter distribution (i.e., $B_\mathrm{e}^\mathrm{NL} \propto B_\mathrm{m}^\mathrm{NL}$).  Refs.~\cite{Hill2016,Ferraro2016,Kusiak_2021} used the bispectrum fitting function from \cite{Gil_Mar_n_2012} (derived
from dark-matter-only N-body simulations) to model $B_\mathrm{m}^\mathrm{NL}$.  An initial comparison of this approach was performed in Ref.~\cite{Ferraro2016} using the hydrodynamical simulations of~\cite{2010ApJ...725...91B}, but the precision was limited by the small volume of the simulations, and in addition the LSS tracer considered in the validation (weak lensing convergence) differed from that used in the data analysis (\emph{WISE} galaxies).  More recently, Ref.~\cite{Bolliet_2023} introduced a halo model with which to compute projected-field kSZ cross-correlations.  In this approach, which we also adopt here, the same approximation is made regarding the dominant term in the projected-field bispectrum, but $B_\mathrm{e}^\mathrm{NL}$ is computed in the halo model, rather than using a fitting function; in addition, no assumption is made that the electrons perfectly trace the dark matter.  Ref.~\cite{Bolliet_2023} used this modeling approach to forecast the projected-field kSZ signal-to-noise ratio for combinations of current and upcoming experiments, including the Atacama Cosmology Telescope (ACT) and Dark Energy Survey (DES), as well as computed cross-correlation predictions with galaxy weak lensing and CMB lensing maps.

In this paper, we compare the halo-model implementation of projected-field kSZ cross-correlations in \verb|class_sz| \cite{Bolliet_2023, bolliet2023classszioverview, bolliet2025classsziinotesexamples} with measurements from the Websky simulations \cite{Websky_2020}.  We find generally good agreement between the theory and simulations, but in some cases differences are found at the level of 50-100$\%$.  In this context, we note that Ref.~\cite{Patki_2023} recently extended the theory calculations of the projected-field kSZ signal to include other terms arising from the Wick contractions of the $\langle TTg\rangle$ bispectrum, i.e., they included all of the non-zero contractions of the $\langle \delta_e v \delta_e v \delta_g \rangle$ five-point function.  They found that the sub-leading terms contribute at the $\sim$10\% level.  However, they considered only the ``effective approach'' for modeling the matter bispectrum, rather than the halo model, which remains to be considered in future work.  In this paper, we wish to assess the validity of the halo model against simulations, but we caution that our halo model in \verb|class_sz| currently implements only the most dominant contraction, the $\langle vv\rangle\langle \delta\delta\delta_g\rangle$ term.  Thus, based on the results of~\cite{Patki_2023}, it is expected that differences on the 10\% level may be present between our theory and the simulations, which indeed is what we find.

This paper is organized as follows. In Sec.~\ref{sec:theory}, we first lay out the general formalism behind LSS cross-correlations, giving the Compton-$y$ and $\tau$ fields as primary examples.  %We introduce the Halo Occupation Distribution (HOD) formalism which describes halos as our LSS tracer in this portion as well. 
We then present the projected-field kSZ framework and the method we use to compute the bispectrum for the halo model.  We then discuss the methodology of our work in Sec.~\ref{sec:sims}.  We detail the Websky simulations and the methods we use to appropriately compare our model with them.  This includes proper considerations of the finite resolution of the simulations, verification of the GNFW \cite{Navarro_1996,Zhao1996} profiles in the field maps, and the construction of our own full-sky temperature maps.  Sec.~\ref{sec:results} presents our main results.  We first present the preliminary validations of our model, showcasing tSZ and $\tau$ cross-correlation angular power spectra for halo samples across different redshifts, before going into the main projected-field kSZ results in which we present a detailed measurement of the kSZ$^2\times$halos signal from Websky in comparison with \verb|class_sz|.  Here, we discuss noise considerations with the halo sample size across redshifts and filters, as well as provide similar statistical calculations \cite{Patki_2023} for comparison.  In Sec.~\ref{sec:discussion}, we discuss future improvements in our model, including adding the other contractions to the five-point function, assessing similarities to the results of Ref.~\cite{Patki_2023}, and forecasting for future surveys. In addition to our primary results, we perform the same projected-field kSZ calculations with a different gas density profile (different experiment filter) for further comparison in Appendix~\ref{Appendix} (Appendix~\ref{Appendix_b}).

%%%%%%%%%%%%%%%%%%%%%%%%%%%%%%%%%%%%%%%%%%%%%%%%%%%%%%%%%%%%%%%%%%%%

\section{Theory: Halo Model}\label{sec:theory}
In this section, we present the halo model formalism we use to calculate angular power spectra of the observables of interest.  The halo model \cite{2000MNRAS.318..203S,COORAY_2002} assumes that all matter in the universe is contained in gravitationally bound halos, which are dominated by dark matter, generally assumed to be spherical in shape, and with constituents distributed in an isotropic way.  These dark matter halos are responsible for the formation of LSS and hence the clustering of galaxies and other LSS tracers.  Here, we first discuss general cross-correlations and then give prescriptions for the Compton-$y$ (tSZ) and $\tau$ fields in the halo model. Finally, we describe the projected-field kSZ signal, the primary observable of interest, and its halo-model implementation.

\subsection{General Cross-Correlation of Two Fields}
Given two tracer fields, their cross-power spectrum can be written as the sum of one- and two-halo terms:
\begin{equation}
    C_{\ell}^{ij} = C_{\ell}^{ij,1h} + C_{\ell}^{ij,2h} \,.
\end{equation}
The one-halo term of the cross-power spectrum between tracers $i$ and $j$ is given by
\begin{equation} \label{1-halo_term}
C_{\ell}^{ij,1h} = \int dz\frac{d^2V}{dzd\Omega}\int dM\frac{dn}{dM}U_{\ell}^i(M,z)U_{\ell}^j(M,z),
\end{equation}
where $dV = \chi^2d\chi$ is the differential comoving volume as a function of redshift, $d\Omega$ is the solid angle, and $\frac{dn}{dM}$ is the halo mass function, which describes the differential number of halos per unit comoving volume per unit mass.  We adopt the Tinker mass function in this work~\cite{Tinker_2008}.  The quantity $U_{\ell}^i(M,z)$ is the multipole-space kernel of a tracer $i$, which we describe further below.

The two-halo term is
\begin{equation} \label{2-halo_term}
C_{\ell}^{ij,2h} = \int dz\frac{d^2V}{dzd\Omega}\int dM_i\frac{dn}{dM_i}b(M_i,z)U_{\ell}^i(M_i,z)\int dM_j\frac{dn}{dM_j}b(M_j,z)U_{\ell}^j(M_j,z)P_{\rm lin}\left(\frac{\ell + 1/2}{\chi},z\right),
\end{equation}
where $b(M,z)$ is the linear halo bias, for which we use the prescription from \cite{2010ApJ...724..878T}, and $P_{\rm lin}\left(\frac{\ell + 1/2}{\chi},z\right)$ is the linear matter power spectrum computed using \verb|class| \cite{lesgourgues2011cosmiclinearanisotropysolving}.

Generally, $U^i_{\ell}(M,z)$ is of the form
\begin{equation} \label{multipole_kernel}
U_{\ell}^i(M,z) = W^i(z) \hat{u}_{\ell}^i(M,z),
\end{equation}
where $W^i(z)$ is the projection kernel that comes from projecting along the LOS and $u_{\ell}^i(M ,z)$ is the Fourier transform of the field's profile for a halo of mass $M$ at redshift $z$. Below, we provide expressions for the fields we consider in this work.

All numerical settings and integration limits in our calculations generally match those in~\cite{Bolliet_2023}.  In particular, all halo model mass integrals span from $10^{10}~M_{\odot}/h$ to $3.5 \times 10^{15}~M_{\odot}/h$ and from $z=0.005$ to 3 (the lower redshift limit avoids numerical instabilities in some fitting functions at $z=0$).  As described further in Sec.~\ref{sec:sims}, we impose an outer radial cutoff of $r_{\rm cut} = 3r_{200c}$, except where noted otherwise.  Details of the full numerical implementation of the halo model calculations can be found in Ref.~\cite{Bolliet_2023}.

\subsubsection{Halo Occupation Distribution}
In our work, we use halos as our LSS tracer. We model them in the halo occupation distribution (HOD) framework \cite{Zheng_2009}, assuming central galaxies only. A halo of a given mass can have at most one central galaxy. The expectation value of the number of central galaxies is dependent on the mass of the halo:
\begin{equation} \label{n_cent}
N_{\mathrm{cent}}(M) = \frac{1}{2}\left(1+\mathrm{erf}\left[\frac{\log_{10}(M/M_{\mathrm{min}})}{\sigma_{\log_{10}M}}\right]\right),
\end{equation}
where $M$ is the halo mass, $M_{\rm min}$ is the characteristic minimum mass of a halo that hosts a central galaxy, and $\sigma_{\log_{10}M}$ characterizes the steepness of the transition between a halo having a central galaxy or not. For this study, we choose $\sigma_{\log_{10}M} = 10^{-7}$ in order to have an abrupt transition in the expectation value. Since our LSS tracer is strictly the halo overdensity field, every halo above the mass threshold should have a central galaxy. Central galaxies trace the center of each halo and are point-like, meaning their real-space density profile is given by the Dirac delta-function. The Fourier transform of our tracer profile $\hat{u}^h_{k}$, with $h$ denoting the halo overdensity, is given by
\begin{equation} \label{halo_fourier}
\hat{u}^h_{k} = \frac{1}{\bar{n}_h}N_{\mathrm{cent}},
\end{equation}
where $\bar{n}_h$ is the mean number of halos per steradian. Lastly, the projection kernel is
\begin{equation} \label{halo_kernel}
W^{\mathrm{\delta_h}}(\chi) = \frac{H}{\chi^2c}\varphi^{\prime}_h(z),
\end{equation}
with 
\begin{equation} \label{normalized_dndz}
\varphi^{\prime}_h(z) = \frac{1}{N_h^{\mathrm{tot}}}\frac{dN_h}{dz},
\end{equation}
and
\begin{equation} \label{num_halos}
N_h^{\mathrm{tot}} = \int dz \frac{dN_h}{dz} \,.
\end{equation}
In these expressions, $H(z)$ is the Hubble parameter and $\frac{dN_h}{dz}$ is the redshift distribution of galaxies, with $\varphi^{\prime}_h(z)$ the normalized redshift distribution.

\subsubsection{Thermal SZ Effect}

\begin{table}[t]
\begin{tabular}{ c | c c c }
  & $A_0$ & $A_M$ & $A_z$ \\
  \hline
 $P_0$ & $18.1$ & $0.154$ & $-0.758$ \\  
 $x_c$ & $0.497$ & $-0.00865$ & $0.731$ \\   
 $\beta$ & $4.35$ & $0.0393$ & $0.415$
\end{tabular}
    \caption{Parameters for electron pressure profile from B12 \cite{Battaglia_2012}, corresponding to their ``AGN feedback'' hydrodynamical simulations~\cite{2010ApJ...725...91B}.}
    \label{B12_params}
\end{table}

The tSZ effect is the temperature shift of the CMB due to the inverse-Compton scattering of CMB photons off of the hot electron gas present in galaxies and clusters.  At frequency $\nu$, the shift in the CMB temperature due to the tSZ effect at angular separation $\vec{\theta}$ on the sky from the center of a halo is given by
\begin{equation} \label{tSZ_field}
\frac{\Delta T^{\rm tSZ}_\nu (\vec{\theta}, M, z)}{T_{\mathrm{CMB}}} = g_{\nu}y(\vec{\theta}, M, z), 
\end{equation}
where $T_{\rm CMB}$ is the mean CMB temperature, $\Delta T^{\mathrm{tSZ}}_\nu$ is the temperature shift of the CMB, and $g_\nu$ is the tSZ spectral energy distribution given by
\begin{equation} \label{spectral_energy_distribution}
g_\nu = q\coth(q/2)-4,
\end{equation}
with $q$ given by
\begin{equation} \label{q_tSZ}
q = \frac{h\nu}{k_{\mathrm{B}}T_{\mathrm{CMB}}},
\end{equation}
where $h$ is Planck's constant and $k_B$ is Boltzmann's constant. Here, $y$ is the Compton-$y$ field,
\begin{equation} \label{compton_y_field}
y(\vec{\theta}, M, z) = \frac{\sigma_{T}}{m_ec^2}\int_{\mathrm{LOS}}P_e\left(\sqrt{l^2 + d_A^2|\vec{\theta}|^2}, M, z\right)dl,
\end{equation}
where $m_e$ is the electron mass, $c$ is the speed of light, $P_e$ is the electron pressure profile, $d_A$ is the angular diameter distance to redshift $z$, and $\sigma_T$ is the Thomson scattering cross-section. The multipole-space kernel $U_{\ell}^y(m,z)$ under the flat-sky approximation for the Compton-$y$ field is~\cite[e.g.,][]{2013PhRvD..88f3526H}
\begin{equation} \label{tSZ_kernel}
U_{\ell}^y(M,z) \approx \frac{4\pi r_s\sigma_T}{\ell_s^2m_ec^2} \int_0^{x_{\rm cut}} dx \, x^2\frac{\sin((\ell + 1/2)x/\ell_s)}{(\ell + 1/2)x/\ell_s}P_e\left(x, M, z\right),
\end{equation}
where $r_s$ is the characteristic scale radius of the pressure profile, $\ell_s$ is the characteristic multipole given by
\begin{equation} \label{ell_s}
\ell_s=\frac{a(z)\chi(z)}{r_s},
\end{equation}
and finally the integration variable $x$ is written as
\begin{equation} \label{tSZ_integration_variable}
x = \frac{a(z)r}{r_s},
\end{equation}
where $r$ is the halo-centric radius of the profile and $a(z)$ is the cosmic scale factor. This formalism assumes that the electron pressure profile is spherically symmetric, which is the case in our model as the thermal pressure profile follows a generalized Navarro–Frenk–White (GNFW) \cite{Navarro_1996,Zhao1996} profile given by
\begin{equation} \label{thermal_profile}
P_{\mathrm{th}}(r) = \frac{P_0 \left(\frac{r}{x_cr_{200c}} \right)^\gamma}{\left[1+\left( \frac{r}{x_cr_{200c}} \right)^\alpha\right]^\beta}P_{200c} \,,
\end{equation} 
where $P_{200c}$ is the self-similar pressure at $r_{200c}$ where $r_{200c}$ denotes the radius that encloses a mass for which the density is 200 times the critical density at that redshift.  We adopt the gas pressure profile fitting function of~\cite{Battaglia_2012}, in which the gas pressure parameters $P_0$, $x_c$, and $\beta$ are taken to be redshift- and mass-dependent, while $\alpha$ and $\gamma$ are set to 1 and 0.3, respectively.  Note that $x_c$, the core scale length, has mass and redshift dependence here, in contrast to the electron density fitting function used below. These parameters are dependent on the halo mass and redshift in the following manner:
\begin{equation} \label{gas_parameter_form}
p(M_{200c},z) = A_0\left(\frac{M_{200c}}{10^{14}M_\odot}\right)^{A_M}(1+z)^{A_z} \,,
\end{equation}
where $p$ is representative of the aforementioned gas parameters. Table \ref{B12_params} contains the exact parameters used in our calculations, which are the best-fit parameters for $A_0$, $A_M$, and $A_z$ from \cite{Battaglia_2012} (hereafter B12).  The electron pressure is directly proportional to the thermal gas pressure via
\begin{equation} \label{pressure_profile_relation}
P_{\mathrm{th}} = \frac{5X_H+3}{2(X_H+1)}P_e \,,
\end{equation}
where $X_H$ is the primordial hydrogen mass fraction, taken to be $X_H=0.76$.  Noting that the projection kernel for the tSZ signal is simply $W^y(\chi) = a(\chi)$~\cite{2013PhRvD..88f3526H}, we have
\begin{equation}
    \hat{u}^y_k = \frac{U_k^y(M,z)}{a(z)}\,,
\end{equation}
where $k$ and $\ell$ are related by $k = (\ell+1/2)/\chi$.  Using Equations~\eqref{1-halo_term}, \eqref{2-halo_term}, \eqref{halo_fourier}, and \eqref{tSZ_kernel}, alongside the given profiles, we calculate the full tSZ-halo cross-correlation.

\subsubsection{Optical Depth}

The optical depth $\tau$ projected along the LOS is defined by
\begin{equation} \label{tau_field}
\tau = \sigma_T \int_{\rm LOS} n_e \, dl \,,
\end{equation}
where $n_e$ is the electron number density. Like $P_{\mathrm{th}}$, we adopt a GNFW profile for $n_e$, which takes the following form:
\begin{equation} \label{electron_num_density}
n_e(r) = \frac{\rho_{\mathrm{gas,free}}(r)}{m_u \mu_e} \,,
\end{equation}
where $m_u$ is the atomic mass unit, $\mu_e \simeq 1.14$ is the mean molecular weight per electron, and
\begin{equation} \label{gas_profile}
\rho_{\mathrm{gas,free}}(r) = f_{\mathrm{b}}f_{\mathrm{free}}\rho_{\mathrm{crit}}(z)C\left(\frac{r}{x_cr_{200c}}\right)^\gamma\left(1+\left[\frac{r}{x_cr_{200c}}\right]^\alpha\right)^{-\frac{\beta + \gamma}{\alpha}} \,,
\end{equation}
where $f_b$ and $f_{\rm free}$ denote the baryon fraction and free electron fraction, respectively.\footnote{In \cite{Battaglia_2016} there is a typo in the gas density profile expression, namely, in the second exponent (N.~Battaglia, priv.~comm.). As opposed to $-\frac{\beta+\gamma}{\alpha}$, it reads $-\frac{\beta-\gamma}{\alpha}$. The correct expression is implemented in \texttt{class\_sz}.}  Note that the parameters $\alpha$, $\beta$, $\gamma$, and $x_c$ are different from those in the gas pressure profile described above, but we maintain the same notation for simplicity.  In the gas density profile, $x_c$ and $\gamma$ are fixed at 0.5 and $-0.2$, respectively. In Table~\ref{B16_params}, we provide the parameter values from \cite{Battaglia_2016} (hereafter B16) that we use for the gas density profile, following the same mass- and redshift-dependence scheme as in Equation~\eqref{gas_parameter_form}.  The Fourier-space optical-depth profile is then given by~\cite{Bolliet_2023}
\begin{equation}
    \hat{u}_k^e = 4\pi \int_0^{r_{\rm cut}} dr \, r^2 \frac{\sin(kr)}{kr} \frac{\rho_{\rm gas, free}(r)}{\bar{\rho}_{m,0}} \,.
\end{equation}

\begin{table}
\begin{tabular}{ c | c c c }
  & $A_0$ & $A_M$ & $A_z$ \\ 
  \hline
 $C$ & $4\times10^3$ & $0.29$ & $-0.66$ \\  
 $\alpha$ & $0.88$ & $-0.03$ & $0.19$ \\   
 $\beta$ & $3.83$ & $0.04$ & $-0.025$
\end{tabular}
\caption{Parameters for electron density profile from B16~\cite{Battaglia_2016}, corresponding to their ``AGN feedback'' hydrodynamical simulations.}
\label{B16_params}
\end{table}

%%%%%%%%%%%%%%%%%%%%%%%%%%%%%%%%%%%%%%%%%%%%%%%%%%%%%%%%%%%%%%%%%%%%%%

\subsection{Projected-Field kSZ Cross-Correlations}
The temperature fluctuations sourced by the kSZ effect at angular separation $\vec{\theta}$ on the sky from the center of a halo are given by
\begin{equation} \label{kSZ_field}
\frac{\Delta T^{\mathrm{kSZ}}(\vec{\theta}, M, z)}{T_{\mathrm{CMB}}} \equiv \Theta^{\mathrm{kSZ}}(\boldsymbol{\hat{n}}) = -\frac{1}{c}\int_{\mathrm{LOS}}d\chi \, g(\boldsymbol{x})\boldsymbol{\hat{n}} \cdot \boldsymbol{v_{e}}(\boldsymbol{x}) \,,
\end{equation}
where $\boldsymbol{\hat{n}}$ denotes the LOS direction, $\boldsymbol{v_{e}}$ is the electron velocity field, and $g(\boldsymbol{x})$ is the visibility function, which characterizes the probability of scattering between CMB photons and electrons within a comoving distance $d\chi$
\begin{equation} \label{visibility_function}
g(\boldsymbol{x}) = a \sigma_T n_e e^{-\tau} \,.
\end{equation}

In the projected-field kSZ estimator, we cross-correlate the square of the filtered kSZ field with an LSS field.  The squaring operation is necessary because the two-point kSZ-LSS cross-correlation vanishes due to the electrons being equally likely to move away from or towards the observer \cite{Dore2004}. Since we are considering cross-correlations between the square of the kSZ field and halos, and considering that the transverse component of the density-modulated electron velocity field $\boldsymbol{\hat{\tilde{v}}_e}^\perp$ sources the kSZ effect, the Fourier-space correlation we are interested in is~\cite{Bolliet_2023}
\begin{equation} \label{three-point_function}
 \langle \boldsymbol{\hat{\tilde{v}}_e}^\perp(\boldsymbol{k})\boldsymbol{\hat{\tilde{v}}_e}^\perp(\boldsymbol{k'})\hat{\delta}_h(\boldsymbol{k''})\rangle = (2\pi)^3\delta_D(\boldsymbol{k}+\boldsymbol{k'}+\boldsymbol{k''}) \mathcal{C}_{v_e^2h}(\boldsymbol{k}, \boldsymbol{k'}, \boldsymbol{k''}) \,,
\end{equation}
with $\boldsymbol{\hat{\tilde{v}}_e}^\perp \sim \delta_e v$ where $\delta_e$ is the electron overdensity field. This makes the three-point function a contraction of the following five-point function (written schematically here):
\begin{gather*} \label{5-point_func_x}
 \langle \delta_e v \delta_e v \delta_h \rangle \sim \langle vv \rangle \langle \delta_e\delta_e\delta_h \rangle
 + \langle \delta_e v \rangle \langle \delta_e v \delta_h \rangle + \langle \delta_e\delta_e \rangle \langle vv\delta_h \rangle \\ + 
 \langle v\delta_e \rangle \langle v\delta_e\delta_h \rangle + 
 \langle v\delta_e \rangle \langle \delta_e v\delta_h \rangle + 
 \langle \delta_e v \rangle \langle v\delta_e\delta_h \rangle \\ + 
 \langle v\delta_h \rangle \langle \delta_e\delta_e v \rangle + 
 \langle \delta_hv \rangle \langle \delta_e\delta_e v \rangle + 
 \langle \delta_e\delta_h \rangle \langle vv\delta_e \rangle + 
 \langle \delta_e\delta_h \rangle \langle vv\delta_e \rangle,
\end{gather*}
with $\delta_h$ being the halo (tracer) density perturbations. From now on, we denote the tracer field specifically as the halo overdensity field. Only four of the Wick contractions above have non-zero contributions at leading order~\cite{DeDeo,Patki_2023}: $\langle vv\rangle\langle \delta_e\delta_e\delta_{h}\rangle$,  $\langle v\delta_e \rangle \langle \delta_e v\delta_h \rangle$, $\langle \delta_e v \rangle \langle v\delta_e \delta_h \rangle$, and $\langle \delta_e\delta_e \rangle \langle vv\delta_h \rangle$.  The most dominant term is $\langle vv \rangle \langle \delta_e\delta_e\delta_h \rangle$, due to the much weaker non-Gaussianity of the velocity field than the small-scale density field~\cite{Dore2004,DeDeo}.  This was confirmed by the initial simulation measurements in~\cite{Ferraro2016}.  Considering only the dominant term, we then have
\begin{equation}\label{bispec_approx}
\mathcal{C}_{v_e^2h} \approx \langle vv \rangle \langle \delta_e\delta_e\delta_h\rangle \approx \frac{2\sigma_v^2}{(f_{\mathrm{b}}f_{\mathrm{free}})^2} B_{\delta_e \delta_e h} \,,
\end{equation}
where $\sigma_v^2 = (1/3) v_{\rm rms}^2/c^2$ is the volume-weighted velocity dispersion and $B_{\delta_e \delta_e h}$ is the bispectrum of two electron density fluctuations and one halo density fluctuation.  A first test of the validity of these approximations was performed using numerical simulations in Refs.~\cite{Hill2016,Ferraro2016}.  However, due to the limited volume of the simulations, the precision of the test was only sufficient for assessing the accuracy of the theoretical model at \emph{Planck} sensitivity.  In this work, we seek to test the theoretical assumptions at much higher precision, with an eye to upcoming high-sensitivity CMB experiments.

In the projected-field estimator we wish to obtain $C_{\ell}^{\mathrm{kSZ^2}\delta_h}$ given by
\begin{equation} \label{projected-fields_estimator}
\langle\Theta_f^2(\ell)\delta_h(\ell')\rangle=(2\pi)^2\delta_D(\ell+\ell')C_{\ell}^{\mathrm{kSZ^2}\delta_h}
\end{equation}
where $\Theta_f$ denotes the Wiener-filtered kSZ field. Before squaring the kSZ field in real space and cross-correlating with the halo overdensity field in harmonic space, we apply a Wiener filter to the kSZ field in order to maximize the signal-to-noise. Combining Equations~\eqref{projected-fields_estimator}, \eqref{three-point_function}, and \eqref{bispec_approx}, the projected-field kSZ power spectrum is thus~\cite{Bolliet_2023}
\begin{equation} \label{projected-fields_cross_correlation}
C_{\ell}^{\mathrm{kSZ^2\delta_h}} = \int_{}\chi^2d\chi W^{\mathrm{kSZ}}(\chi)^2W^{\mathrm{\delta_h}}(\chi)T(\ell,\chi) \,,
\end{equation}
where the kSZ projection kernel $W^{\mathrm{kSZ}}(\chi)$ is
\begin{equation} \label{kSZ_kernel}
W^{\mathrm{kSZ}}(\chi) = \frac{a \sigma_T \bar{\rho}_m \sigma_v}{m_\mu \mu_e \chi^2} \,,
\end{equation}
$\bar{\rho}_m$ is the mean matter density, and $W^{\delta_h}(\chi)$ is again the same halo overdensity projection kernel in Equation~\eqref{halo_kernel}. The ``triangle power spectrum'' $T(\ell,\chi)$ is
\begin{equation} \label{triangle_power_spectrum}
T(\ell,\chi) = \int_{}\frac{d^2\boldsymbol{\ell '}}{(2\pi)^2}w(\ell')w(\abs{\boldsymbol{\ell'+\ell}})B_{\delta_e \delta_e h} (\boldsymbol{k_1}, \boldsymbol{k_2}, \boldsymbol{k_3}) \,,
\end{equation}
where $\boldsymbol{\ell}^\prime = \boldsymbol{k}_1 \chi, -(\boldsymbol{\ell} + \boldsymbol{\ell}^\prime) = \boldsymbol{k}_2 \chi$, and $\boldsymbol{\ell} = \boldsymbol{k}_3 \chi$.  The triangle power spectrum integrates over contributions from all triangles with sides given by $(\boldsymbol{\ell}, \boldsymbol{\ell}^\prime, -(\boldsymbol{\ell} + \boldsymbol{\ell}^\prime))$ lying in planes of constant redshift perpendicular to the LOS~\cite{Dore2004}.
Lastly, $w(\ell)$ is the Wiener filter applied to the temperature map to upweight the kSZ signal on relevant scales.  We also define the filter to vanish at low $\ell$ in order to remove any contributions from the Integrated Sachs-Wolf (ISW) effect~\cite{Hill2016,Kusiak_2021}.  In this study, we use filters constructed for \textit{Planck} \cite{LGMCA} and the Simons Observatory (SO) \cite{Ade_2019,SO2025}, which are low- and high-resolution experiments, respectively (SO is also much lower-noise than \emph{Planck}). The Wiener filter is applied to a beam-convolved temperature map; thus, $w(\ell)$ takes the following form:
\begin{equation} \label{wiener_filter}
 w(\ell) = b(\ell)F(\ell) \,,
 \end{equation}
where we take the beam $b(\ell)$ to be a Gaussian,
\begin{equation} \label{gauss_beam}
 b(\ell) = \exp{-\frac{1}{2}\ell(\ell+1)\frac{\theta_{\mathrm{FWHM}}^2}{8\ln{2}}} \,,
\end{equation}
where $\theta_{\mathrm{FWHM}}$ is the full-width at half-maximum of the telescope's beam in radians.  The filter $F(\ell)$ is given by~\cite{Bolliet_2023}:
\begin{equation} \label{filter}
F(\ell) = \frac{\sqrt{C_{\ell}^{\mathrm{kSZ}}}/C_{\ell}^{\mathrm{tot}}}{\mathrm{max}\left(\sqrt{C_{\ell}^{\mathrm{kSZ}}}/C_{\ell}^{\mathrm{tot}}\right)} \,,
\end{equation}
where $C_{\ell}^{\mathrm{kSZ}}$ is the kSZ auto-power spectrum, which we take from the B12 simulations.
% JCH: we don't use this; we use a kSZ power spectrum from simulations
%given by 
%\begin{equation} \label{cl_ksz}
%C_{\ell}^{\mathrm{kSZ}} = \int_{}\chi^2d\chi W^{\mathrm{kSZ}}(\chi)^2P_{\delta_e\delta_e}(k,\chi),
%\end{equation}
%which is the high-k limit form of the ksz auto power spectrum where $P_{\delta_e\delta_e}$ is the electron power spectrum, computed from hydrodynamical simulations. 
Note that $C_{\ell}^{\mathrm{tot}}$ is the total auto-power spectrum of the temperature map, containing CMB, kSZ, noise, and post-component separation foregrounds, such as the cosmic infrared background:
\begin{equation} \label{cl_tot}
C_{\ell}^{\mathrm{tot}} = C_{\ell}^{\mathrm{kSZ}} + C_{\ell}^{\mathrm{\Theta\Theta,lensed}} + N_{\ell}^{\Theta\Theta} \,,
\end{equation}
highlighting the lensed primary CMB and noise. In this paper, we compare the theory with simulated maps that only contain the kSZ signal, so we do not model residual foregrounds in the signal calculation, but we do include the effects of post-component-separation foregrounds in the noise, i.e., in $N_{\ell}^{\Theta\Theta}$.  Note that \textit{Planck} and SO have beam FWHMs of $5$ and $1.4$ arcminutes, respectively.  By working with beam-convolved signal maps, we avoid numerical divergences at small scales.  The \emph{Planck} and SO filters are shown in Fig.~\ref{filters_plot}.

\begin{figure}[t]
    \centering
    \includegraphics[width=0.7\columnwidth]{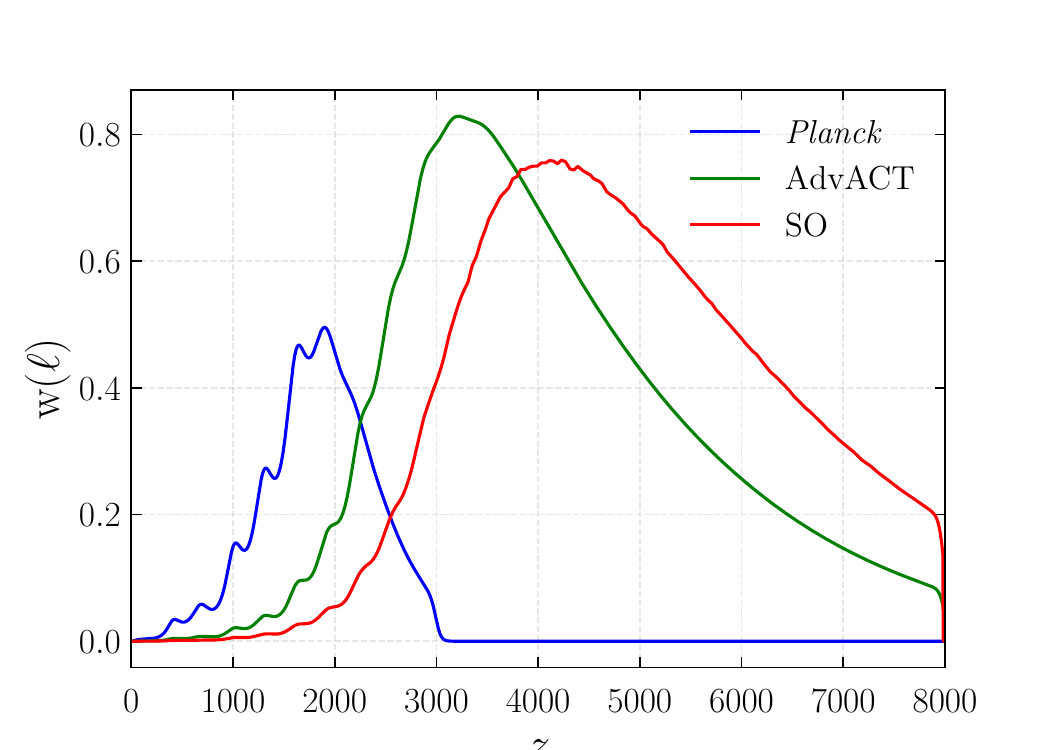}
    \caption{ \textit{Planck} (blue), AdvACT (green), and SO (red) beam-convolved Wiener filters, as applied to our kSZ maps. The \textit{Planck} filter truncates to zero at $\ell \approx3000$ and the  AdvACT and SO filters at $\ell\approx8000$. All three filters remove very low-$\ell$ modes to avoid contributions from the ISW effect.}
    \label{filters_plot}
\end{figure}

There are two primary ways of modeling $B_{\delta_e \delta_e h}$: the ``effective approach'' adopted in~\cite{Hill2016,Ferraro2016,Kusiak_2021} and the halo model approach developed in~\cite{Bolliet_2023}.  We primarily use the halo model in this work, but we briefly discuss the effective approach as well.  In the halo model, the hybrid bispectrum can be written as the sum of the one-halo, two-halo, and three-halo terms:
\begin{equation} \label{bispectrum}
B_{\delta_e \delta_e \mathrm{h}} = B_{\delta_e \delta_e \mathrm{h}}^{1\mathrm{h}} + B_{\delta_e \delta_e \mathrm{h}}^{2\mathrm{h}} +B_{\delta_e \delta_e \mathrm{h}}^{3\mathrm{h}} \,.
\end{equation}
These are respectively given by (note that we suppress the redshift dependence of all quantities here for brevity):
\begin{equation} \label{1_halo_bispec}
 B_{\delta_e \delta_e \mathrm{h}}^{1\mathrm{h}} = \int_{} dM \frac{dn}{dM} \hat{u}_{k_1}^e(M)\hat{u}_{k_2}^e(M)\hat{u}_{k_3}^{\mathrm{h}}(M),
\end{equation}

\begin{equation} \label{2_halo_bispec}
 B_{\delta_e \delta_e \mathrm{h}}^{2\mathrm{h}} = \int_{}dM_1\frac{dn}{dM_1}b(M_1)\hat{u}_{k_1}^e(M_1)\hat{u}_{k_2}^e(M_1)\int_{}dM_2\frac{dn}{dM_2}b(M_2)\hat{u}_{k_3}^{\mathrm{h}}(M_2)P_{\mathrm{lin}}(k_3) + \mathrm{perms.} , 
\end{equation}

\begin{align}
\label{3_halo_bispec}
 B_{\delta_e \delta_e \mathrm{h}}^{3\mathrm{h}} = 2\int_{}dM_1\frac{dn}{dM_1}b(M_1)\hat{u}_{k_1}^e(M_1)P_{\mathrm{lin}}(k_1) \int_{}dM_2\frac{dn}{dM_2}b(M_2)\hat{u}_{k_2}^e(M_2)P_{\mathrm{lin}}(k_2) \int_{}dM_3\frac{dn}{dM_3}b(M_3)\hat{u}_{k_3}^{\mathrm{h}}(M_3) F_2(k_1,k_2,k_3) \nonumber \\
     + \int dM_1 \frac{dn}{dM_1} b(M_1) \hat{u}_{k_1}^e(M_1) P_{\mathrm{lin}}(k_1) \int dM_2 \frac{dn}{dM_2} b(M_2) \hat{u}_{k_2}^e(M_2) P_{\mathrm{lin}}(k_2) \int dM_3 \frac{dn}{dM_3} b^{(2)}(M_3) \hat{u}_{k_3}^{\mathrm{h}}(M_3) + \mathrm{perms.} ,
\end{align}
where $F_2(k_1,k_2,k_3)$ is the coupling kernel that appears in the tree-level bispectrum in perturbation theory and $b^{(2)}(M,z)$ is the second-order halo bias.  Further details can be found in Ref.~\cite{Bolliet_2023}.

In the ``effective approach'', as recently detailed in \cite{Patki_2023} and first introduced in \cite{scoccimarro2001}, the hybrid bispectrum is modeled via fitting functions from N-body simulations.  This leads to the following form:
\begin{equation} \label{bispec_eff}
B_{\delta_e \delta_e \mathrm{h}}^{\mathrm{eff}} = f_b^2 f_{\rm free}^2 b_h \left( 2 F_2^{\rm eff}(k_1,k_2,k_3) P_{\mathrm{non-lin}}(k_1)P_{\mathrm{non-lin}}(k_2) + \mathrm{two \,\, \space perms.} \right),
\end{equation}
where $P_{\mathrm{non-lin}}(k)$ is the non-linear matter power spectrum, $b_h$ is the population-averaged linear bias of the halo sample, and $F_2^{\rm eff}$ is a redshift- and scale-dependent effective kernel, for which we use the fitting function of~\cite{Gil_Mar_n_2012}.  In the effective approach, it is assumed that the electrons trace the overall dark matter distribution, and that the halos (or other LSS tracers) are linearly biased tracers of the dark matter field.  Unlike the halo model, this approach thus cannot easily model any non-gravitational gas physics due to the lack of any specified profiles.  Further details can be found in Refs.~\cite{Dore2004,DeDeo,Hill2016,Ferraro2016,Bolliet_2023,Patki_2023}.

For all halo model calculations, as well as both methods for computing projected-field kSZ cross-correlations (the halo model and the effective approach), we use \verb|class_sz| \cite{Bolliet_2023}, a halo model code capable of efficiently computing power spectra and bispectra for various LSS tracers, including different parameters for HOD prescriptions and GNFW profiles.

\section{Simulations and Methods} \label{sec:sims}
The primary simulation that we analyze in this work is that of Websky~\cite{Websky_2020}.  Websky is a ``peak-patch'' simulation of the formation of cosmological LSS, which follows the gravitational evolution of the dark matter field.  Baryonic observables are ``painted'' onto the simulation in post-processing, generally by using specified profiles that are applied to the dark matter halos.  From these painted halos, Websky generates full-sky \texttt{HEALPix} maps of relevant cosmological fields, including maps of the kSZ, tSZ, and CMB lensing signals.  For the tSZ and kSZ maps, the halos are painted using GNFW profiles for the electron pressure and density, respectively.

We use the Websky halo catalog to build halo density maps, which we then cross-correlate with other fields of interest. The Websky catalog contains halos in the redshift range $0 < z < 4.6$, with most in the lower range of $0 < z < 2.5$.  Within the entire catalog there are roughly $870$ million halos. When comparing our halo-model results to Websky, we consider three different redshift ranges: $0.2 < z < 0.5$, $0.5 < z < 1.0$, and $1.0 < z < 2.5$. However, we consider a fixed mass range of $6 \times 10^{13} \, M_\odot /h < M < 5 \times 10^{15} \, M_\odot /h$ throughout the analysis.  The lower limit of this range is determined by the mass at which the Websky halo catalogs become incomplete, when compared to the predicted number of halos from the Tinker halo mass function~\cite{Tinker_2008}.  Tables~\ref{halo_num_z} and~\ref{websky_cosmology} provide the number of halos in the catalog for each redshift cut and the Websky cosmological parameters, respectively.

\begin{table}[t]
\begin{tabular}{ |c|c|c|}
$0.2<z<0.5$ & $0.5<z<1$ & $1<z<2.5$ \\ 
\hline
332434  & 707105 & 384444 
\end{tabular}
    \caption{
    Number of Websky halos in each redshift cut with masses between $6\times 10^{13} \, M_\odot /h  < M < 5 \times 10^{15} \, M_\odot /h$.}
    \label{halo_num_z}
\end{table}

\begin{table}[t]
\begin{tabular}{ |c|c|c|c|c|c|c|c|c|}
$\Omega_{m}$ & $\Omega_{b}$ & $\Omega_{\Lambda}$ & $h$ & $\sigma_8$ & $n_s$ \\ 
\hline
0.31 & 0.049 & 0.69 & 0.68 & 0.81 & 0.965
\end{tabular}
    \caption{Websky cosmological parameters.}
    \label{websky_cosmology}
\end{table}

We consider both the Websky-produced maps of CMB secondary anisotropy signals and maps that we produce via our own painting algorithm applied to the Websky halo catalogs.  For the tSZ signal, we consider both the Websky Compton-$y$ map and our own painted map.  For the $\tau$ and kSZ signals, we generate our own maps due to difficulties in interpreting the outer radial cutoff of the painted electron density profiles in Websky.

We generate our own tSZ, kSZ, and $\tau$ maps using the lightcone halo catalog from Websky. As described above, these effects are sourced by the projected electron pressure ($P_e$), electron momentum, and electron density ($n_e$), respectively. As mentioned in Sec.~\ref{sec:theory}, we use GNFW models for the halo-centric profiles of these fields, which evolve with the halo mass and redshift. The best-fit parameters for these profiles were obtained by fitting them to hydrodynamical simulations of \cite{2010ApJ...725...91B}.

We follow the general pasting procedure similar to the description in \cite{Websky_2020}. Assuming a 3D radially symmetric profile of the probe of interest, $f_{\rm 3D}(r, M, z)$, for a halo of mass $M$ at redshift $z$, we first estimate the corresponding projected 2D profile, $f_{\rm 2D}(r_p, M, z) = \int_{-R_{\rm max}}^{R_{\rm max}} f_{\rm 3D}(r, M, z) d r_{||}$, by integrating along the physical LOS direction. Here, the integration limits are related to the spherical overdensity radius of each halo, $R_{\rm max} = 3 R_{\rm 200c}(M,z)$. We then project these profiles for all halos in the lightcone onto a \texttt{HEALPix} grid with $N_{\rm side} = 8192$. For this purpose, we first identify the \texttt{HEALPix} pixels that overlap with each halo within a radius $R_{\rm max}$  from its center and evaluate the interpolated projected profiles at those pixel locations. Note that we convert the reported $M_{\rm 200m}$ masses in the Websky halo catalog to $M_{\rm 200c}$ assuming the concentration-mass relation from \cite{Bhattacharya:2013:ApJ:}.

For the tSZ map (Compton-$y$), we have $f_{\rm 3D} = P_e$, and we linearly sum the projected profiles for all the halos, which is then converted to the correct units to obtain $y = \frac{\sigma_T}{m_e c^2} \int P_e dl$. Similarly, for the $\tau$ map, we have $f_{\rm 3D} = n_e$, and by repeating the above procedure, we obtain $\tau = \sigma_T \int n_e dl$. However, for the kSZ map, we additionally have to multiply the projected profile for each halo by its LOS velocity, obtaining $T_{\rm kSZ} = \frac{\sigma_T}{c} \int n_e v_{e, ||} dl$. Note that here, we only account for the halo component of the kSZ signal and ignore the contributions from particles not associated with any halo. While non-halo contributions are important for the auto-correlation of the kSZ field~\cite[e.g.,][]{2018ApJ...853..121P}, which probes the low-density electron distribution, we can safely ignore these contributions to the small-scale cross-correlation statistics that we are interested in here \cite{Bolliet_2023}.

For a proper comparison between the simulations and theoretical prediction, the outer cut-off of the halos must be specified. The total gas mass of a given halo diverges when integrating Equation~\ref{gas_profile}. It is therefore important to specify a hard cut to the size of each halo. It is unclear what exact cut was used in constructing the Websky kSZ map. We therefore resort to pasting B16 electron density profiles with the gas parameters from Table~\ref{B16_params} for the kSZ projected-field calculation. The same is done for the tSZ validation, pasting B12 Compton-$y$ profiles instead, assuming the parameter values from Table~\ref{B12_params}. Additionally, although Websky originally does not produce these maps, we generate $\tau$ maps to enable further diagnosis of our model.  We measure the cross-correlation of the $y$ (or $\tau$) maps with the halo number density maps using \texttt{healpy}, as all maps are full-sky with no mask.

To measure the projected-field kSZ cross-correlations, we must first Wiener-filter the $T_{\rm kSZ}$ maps and then square them in pixel space, and then cross-correlate them with the halo overdensity maps of the appropriate redshift and mass range. To have an accurate comparison between our model and Websky, we must take into account the resolution and box size of the Websky simulations and the maps they produce. The simulations are in a $(15.4~\mathrm{Gpc})^3$ light cone, in $10^{12}$ resolution elements. The box size directly affects the kSZ projection kernel in Equation~\eqref{kSZ_kernel}, via the velocity dispersion $\sigma_v$.  Note that $\sigma_v$ can be written as $\sigma_v^2 = \frac{v_{\mathrm{rms}}^2}{3c^2}$. Analytically, the rms velocity $v_{\rm rms}$ is given by
\begin{equation} \label{v_rms}
v^2_{\mathrm{rms}}(\chi) = \frac{1}{2\pi^2}\int dk \, k^2 P_{vv}(k,\chi) \,,
\end{equation}
where $P_{vv}$ is the velocity power spectrum, computed using the non-linear matter power spectrum (note that the non-linear corrections are small). In reality, Equation~\ref{v_rms} should be integrated over all of $k$-space, but due to the finite resolution and box size of the simulation, there are lower and upper bounds in the integral.  The lower bound is set by the largest mode that the box contains, while the upper bound is set by the Nyquist frequency and/or gravitational softening length of the simulation.  To account for these effects, instead of modifying our calculation of $v_{\rm rms}$, we directly compute the rms velocity of the halos that we select from the Websky halo catalog and use that instead in the theoretical computation. This primarily affects cross-correlations for halos at redshifts $0 < z < 1$, as seen in Fig.~\ref{v_rms_plot}, which compares the rms velocity computed using Equation~\eqref{v_rms} (including approximate corrections for the minimum and maximum wavenumbers) to that directly measured from the Websky halos. At $z > 1$ there is very close agreement, with Websky slightly exceeding the analytic calculation at $z \gtrsim 2$.

\begin{figure}[t]
    \centering
    \includegraphics[width=0.6\columnwidth]{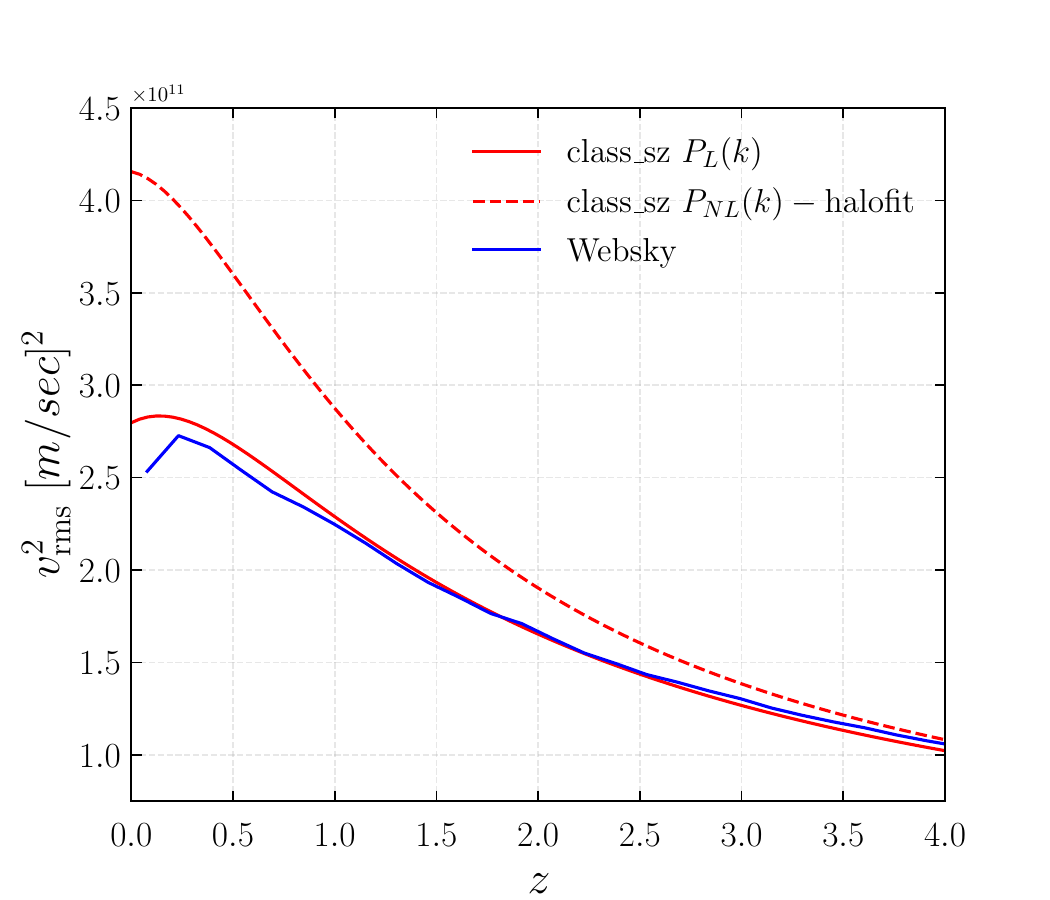}
    \caption{Websky $v_{\rm rms}^2$ (blue) versus \texttt{class\_sz} prediction using Equation~\eqref{v_rms} for the linear (red solid) and non-linear (red dashed) matter power spectrum, the latter computed with the Halofit prescription~\cite{Halofit}. The $v_{\rm rms}^2$ measured from Websky agrees better with the linear-theory prediction.  For the redshifts we are considering in this study there is a notable difference in the Websky and nonlinear theory curves, leading to an overall decrease in the theoretical computation if not properly accounted for. This affects projected-field kSZ cross-correlations with halos at lower redshifts, primarily halos at $0<z<1$. The difference between the Websky and linear-theory curves is at the $\sim 5\%$ level; we use the Websky curve shown here in our \texttt{class\_sz} calculations when comparing to the simulation results.}
     \label{v_rms_plot}
\end{figure}

%%%%%%%%%%%%%%%%%%%%%%%%%%%%%%%%%%%%%%%%%%%%%%%%%%%%%%%%%%%%%

\section{Results}
\label{sec:results}
\subsection{Validation of Simulated tSZ and $\tau$ Maps}
Using our newly constructed tSZ and $\tau$ maps described above, we perform a cross-correlation with the Websky halos.  We use \texttt{HEALPix} maps with resolution $N_{\rm side}=8192$ and measure the cross--spectra with \texttt{healpy} on unmasked, full-sky maps. We bin them in linearly spaced multipole bins of width $\Delta\ell=250$. In Figs.~\ref{tau_correlation} and~\ref{tsz_correlation}, we show the tSZ and $\tau$ cross-correlation results, compared with the \verb|class_sz| halo model prediction.  We emphasize that the cosmological parameters, pressure profiles, and gas density profiles are all set identically in the simulated maps and in \verb|class_sz|.

\begin{figure}[t]
  \centering
  \includegraphics[width=.8\linewidth]{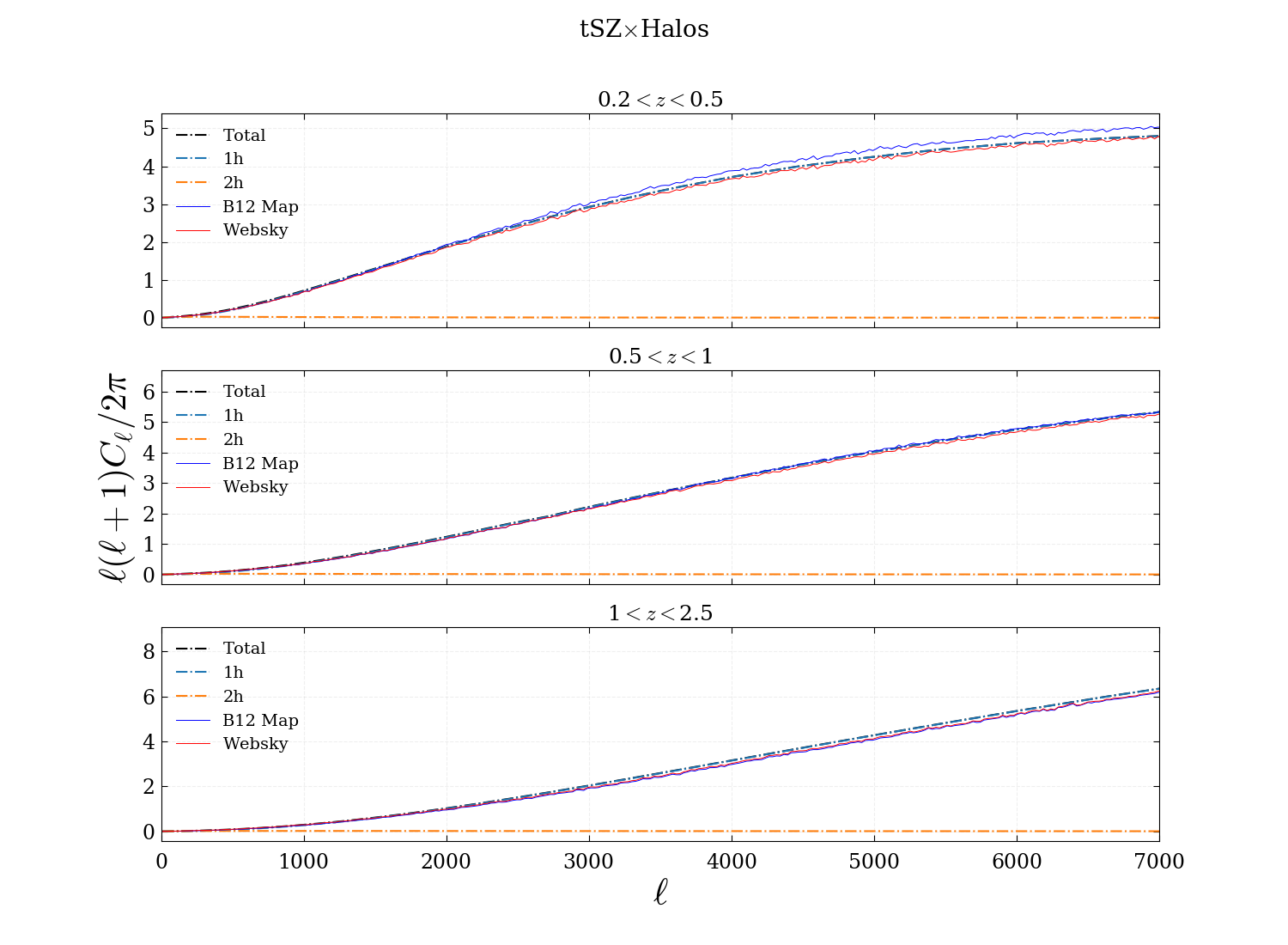}
  \caption{Comparison of tSZ--halo cross-power spectra between theory and simulations for three redshift ranges, as indicated in the plot titles. In the dash-dotted lines, we show the {\tt class\_sz} halo model prediction computed with the B12 pressure profile from Table~\ref{B12_params} (with the one-halo, two-halo, and total spectra shown individually as labeled), and in solid we show the cross-correlation measured from simulations: the Websky Compton-$y$ map in red and our constructed Compton-$y$ maps (see Sec.~\ref{sec:sims}) in blue.  All curves agree to a very good level.  Note that the pressure profile cut-offs for our pasted map and the Websky map are $3R_{200c}$ and $4R_{200c}$, respectively, but this minor difference does not result in drastic differences in the measured cross--spectra.}
  \label{tsz_correlation}
\end{figure}

\begin{figure}[t]
  \centering
\includegraphics[width=.8\linewidth]{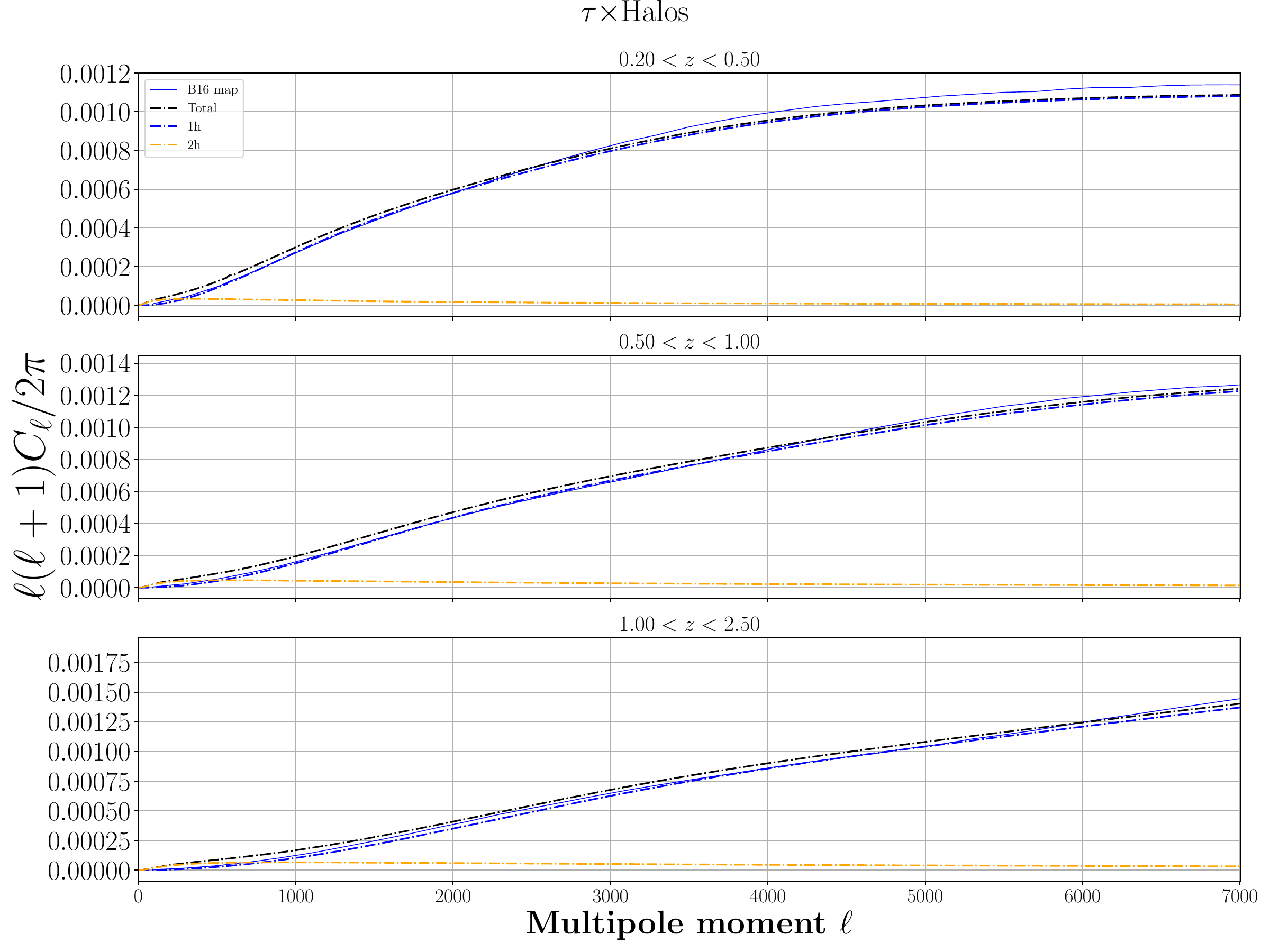}
  \caption{Comparison of $\tau$--halo cross-power spectra between theory and simulations for three redshift ranges, as indicated in the plot titles.
  In the dash-dotted lines, we show the {\tt class\_sz} halo model prediction computed with the B16 gas density profile from Table~\ref{B16_params} (with the one-halo, two-halo, and total spectra shown individually as labeled), and in solid blue we show the cross-correlation measured from simulations using our constructed $\tau$ maps (see Sec.~\ref{sec:sims}). (Note that the Websky suite does not provide $\tau$ maps.)  The overall agreement is very good, with only a few-percent discrepancy present on small scales in the $0.2<z<0.5$ redshift cut.}
  \label{tau_correlation}
\end{figure}

We present the tSZ-halo cross-correlation results in Fig.~\ref{tsz_correlation} for both our pasted B12 Compton-$y$ map and the original Websky Compton-$y$ map. It is not a completely apples-to-apples comparison because the pasted maps truncate the pressure profiles at $3R_{200c}$, whereas Websky truncates the profiles at $4R_{200c}$; our \verb|class_sz| theoretical curves assume the former.  However, despite this difference, the Websky, pasted B12, and \verb|class_sz| curves all agree within 1\%, with \verb|class_sz| slightly favoring the Websky result for the $0.2 < z < 0.5$ redshift cut. For the $0.5 < z < 1$ and $1 < z < 2.5$ cuts, the pasted B12 Compton-$y$ map curves are much closer. For the latter cut, the pasted maps and Websky are practically identical, with \verb|class_sz| slightly over-predicting the simulations by less than 1\%. At roughly $\ell = 3000$, the curves diverge slightly.  Similar behavior can be seen in the $\tau$-halo cross-correlation in Fig.~\ref{tau_correlation}. %, with the additional observation that the simulation and theory curves trend towards convergence at $\ell = 10000$. 
%The $\tau$ results also exhibit a slight oscillatory behavior that the Compton-$y$ maps do not contain. This is especially prevalent in the $1 < z < 2.5$ cuts that \verb|class_sz| does not produce. For example, at roughly $\ell = 3000$ \verb|class_sz|'s prediction is larger the map's. Then at around $\ell = 6000$ \verb|class_sz| dips below the map's angular power spectrum. It turns over at $\ell = 1000$ once more. 
It should be noted that we correct the Websky curves with one power of the pixel window function (to account for the pixel window in the Websky Compton-$y$ map), while the pasted maps are not corrected, as our method to produce these maps does not involve averaging the signal within pixels.

With these results, we confirm that the simulated maps implement the same profiles as used in \verb|class_sz| and that the halo mass range we choose exhibits good agreement with the Tinker halo mass function (note that the tSZ and $\tau$ cross-correlations are sensitive to different halo masses, due to the $\sim M^{5/3}$ scaling of the tSZ signal versus the $\sim M$ scaling of the $\tau$ signal with halo mass). In particular, the $\tau$ cross-correlation agreement is an important validation that the electron density profile in the simulated maps is implemented properly, thus allowing us to proceed with the projected-field kSZ analysis.

\subsection{Projected-Field kSZ Cross-Correlation Results}
\label{subsec:kSZ}
In this subsection, we present the main results of this work, namely, comparisons between halo-model calculations of projected-field kSZ cross-correlations, as implemented in \verb|class_sz| and measurements from simulations. 
We consider three halo samples from Websky spanning different redshift ranges, as described above.  We use the pasted B16 kSZ maps that we construct from the Websky halo catalogs (see Sec.~\ref{sec:sims}), rather than the original Websky kSZ map, due to difficulties in interpreting the profiles in the Websky map.  We apply beam-convolved filters to the kSZ map for \textit{Planck} and SO, as described above, and then square the filtered kSZ maps in real space and cross-correlate them with the halo number density maps.

Our main results are shown in Fig.~\ref{ksz_so} for the SO setup and in Fig.~\ref{ksz_planck} for \emph{Planck}, for the three different redshift cuts. In the left panels, the dash-dotted black curve shows the total halo-model prediction computed with \verb|class_sz|, which is a sum of the one-halo (dashed-dotted blue), two-halo (dashed-dotted orange), and three-halo (dashed-dotted green) terms, with the one-halo term dominating significantly in all cases. The simulation measurements are shown as thin blue curves with error bars (described further below). The red dashed lines show a simple rescaling of the \verb|class_sz| prediction by an overall amplitude $A$ (with values given in the legend and discussed in detail below). The right panels show the ratio of the \verb|class_sz| predictions to the simulations across the three redshift bins.

For the SO case in Fig.~\ref{ksz_so}, the theory and simulation curves agree at the $\sim 50\%$ level, with scale dependence only present at very low $\ell$, i.e., the analytic theory is larger than the simulation results by roughly a constant factor over most of the multipole range. For the \emph{Planck} setup shown in Fig.~\ref{ksz_planck}, the level of agreement is within a factor of $\sim 1.7$-2, with some scale dependence at very low and very high $\ell$ (though the latter is well into the noise-dominated range for \emph{Planck}), but also relatively constant. Our results show a striking lack of any significant scale dependence, especially when compared with the results of~\cite{Patki_2023}.  However, the validation of the tSZ and $\tau$ cross-correlations (Figs.~\ref{tsz_correlation} and~\ref{tau_correlation}), and our use of the $v_\mathrm{rms}$ from the Websky halo catalogs, provide a robust check that the difference does not come from a mismatch of the gas density profiles, halo mass function, or velocities.  We interpret the difference as arising from the sub-leading terms in the halo model calculation (recall that we only include the dominant term contributing to five-point function), as well as the approximations of the halo model itself.  In addition, we note that the magnitude of the fractional difference between our theory and simulation curves is similar to that found by~\cite{Patki_2023} in their comparison of the full theory calculation (including all Wick contractions) to the dominant term (the only term included in our theory calculation, cf.~Equation~\eqref{bispec_approx}).  A more detailed comparison with Ref.~\cite{Patki_2023} is not straightforward, as they used the ``effective approach'' to compute the signal, whereas our theory calculation uses the halo model.  Nevertheless, the order-of-magnitude agreement of the fractional contribution of the missing terms is reassuring.

To more quantitatively assess the accuracy of our halo-based projected-field kSZ model, we compute error bars on the simulation measurements.   We use two approaches: a Monte Carlo–based method and a Gaussian covariance approximation. In the Monte Carlo method, we generate CMB map realizations containing noise consistent with the relevant experimental power spectra.  For SO, we use publicly available post-component-separation noise power spectra~\cite{Ade_2019}\footnote{\url{https://github.com/simonsobs/so_noise_models/tree/master/LAT_comp_sep_noise/v3.1.0}}.  We emphasize that these noise power spectra include the effects of residual foregrounds (at the Gaussian level), which are important on small scales.  For \emph{Planck}, we include only the lensed CMB (and kSZ) in our sky maps, due to the small number of halos in the Websky samples, which leads to large shot noise in the cross-correlation (if we use the ``LGMCA'' noise map power spectra for \emph{Planck}~\cite{LGMCA}, the error bars on the resulting cross-correlation are too large to yield meaningful conclusions}.\footnote{We emphasize that the \emph{Planck} filter used in this work does include the effects of noise, i.e., it isolates the same scales as used in the measurement of Ref.~\cite{Kusiak_2021}.}  To each CMB (+ noise) realization, we add the simulated kSZ map, and then carry out the usual filtering, squaring, and cross-correlation with the halo overdensity maps. This procedure is repeated for 1000 realizations, from which we construct the covariance matrix, obtain error bars, and evaluate the $\chi^2$ for fitting.

\begin{figure}[t]
  \centering
  \includegraphics[width=\linewidth]{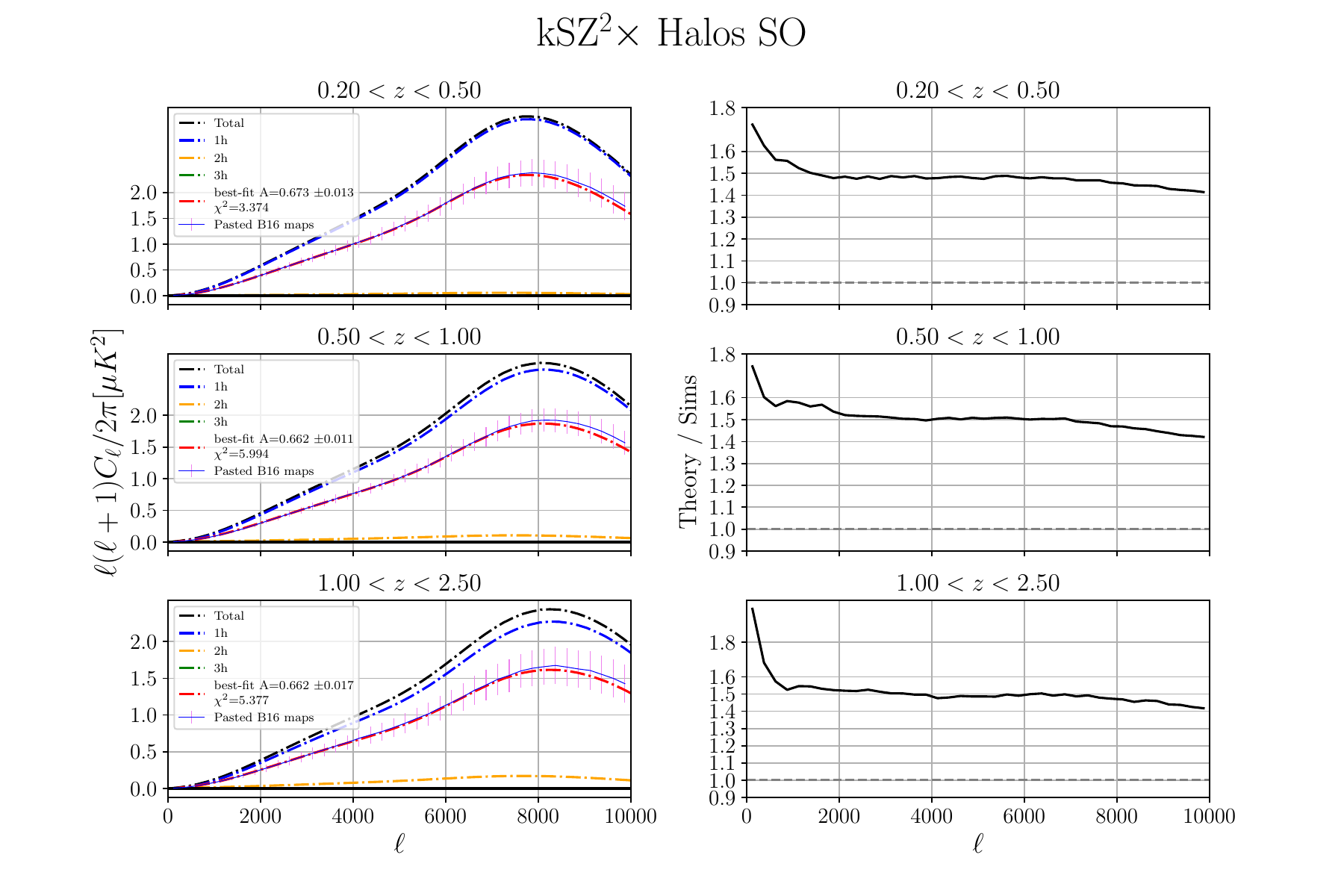}
  \caption{Comparison of kSZ$^2$--halo cross-power spectra between theory and simulations (left) and ratio between theory and simulation curves (right) for three redshift ranges, as indicated in the plot titles.  All computations here are for the SO experimental configuration.  In the left panels, the dash-dotted black lines denote the total halo-model prediction, comprised of the one-halo (dashed-dotted blue), two-halo (dashed-dotted orange), and three-halo (dashed-dotted green) terms. The thin blue lines with error bars show the results measured from simulations, with errors computed via the Gaussian covariance approximation. The dashed red lines indicate a rescaling of the {\tt class\_sz} halo model curve by an overall amplitude $A$ (best-fit value given in the legend) to match the measured simulation result, as obtained via $\chi^2$ fitting.  In the right panels, we show the corresponding theory-to-simulation ratios, which are mostly scale-independent and close to $\approx 1.5$ at $\ell \gtrsim 300$, with the exception of the $0.2<z<0.5$ redshift range yielding a slightly lower ratio.}
  \label{ksz_so}
\end{figure}

\begin{figure}[t]
  \centering
  \includegraphics[width=\linewidth]{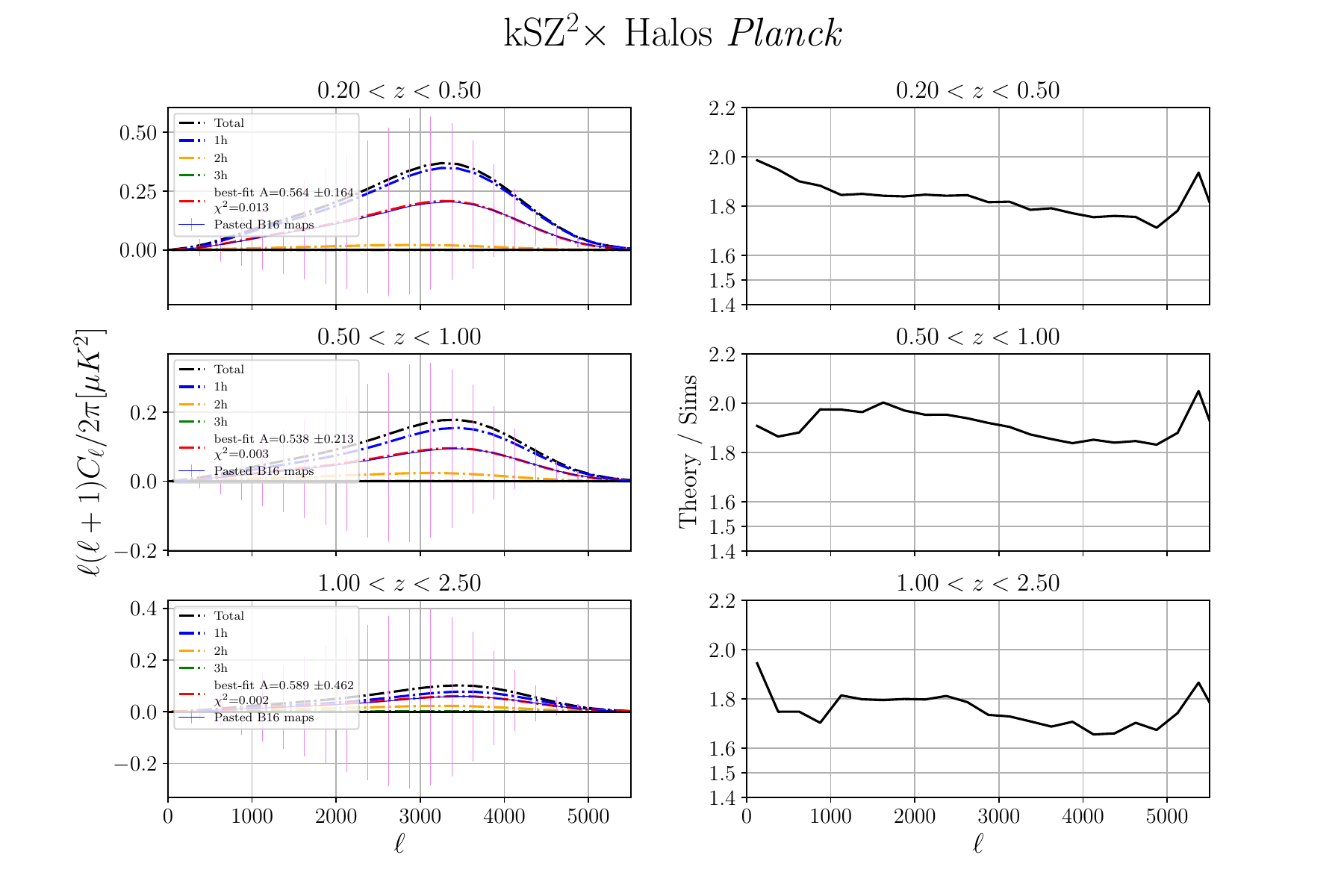}
  \caption{omparison of kSZ$^2$--halo cross-power spectra between theory and simulations (left) and the theory-to-simulation ratio (right) for three redshift ranges, as indicated in the plot titles.  All computations here are for the \emph{Planck} experimental configuration.  In the left panels, the dash-dotted black lines denote the total halo-model prediction, comprised of the one-halo (dashed-dotted blue), two-halo (dashed-dotted orange), and three-halo (dashed-dotted green) terms. The thin blue lines with error bars show the results measured from simulations, with errors computed via the Gaussian covariance approximation. The dashed red lines indicate a rescaling of the {\tt class\_sz} halo model curve by an overall amplitude $A$ (best-fit value given in the legend) to match the measured simulation result, as obtained via $\chi^2$ fitting.  In the right panels, we show the corresponding theory to simulations ratios, which range between $\approx 1.7$-2.0 (note that the very high-$\ell$ features are deep into the noise-dominated regime for \emph{Planck}).}
  \label{ksz_planck}
\end{figure}

Our second approach is to estimate the error bars using a Gaussian covariance approximation, in which the covariance matrix is given by
\begin{equation} \label{gauss_cov}
M_{\ell\ell'}=\frac{1}{(2\ell+1) \Delta\ell} \left\{C_\ell^{\Theta_f^2\Theta_f^2} \left(C_\ell^{\delta_h\delta_h} + N_\ell^{\delta_h\delta_h} \right) + \left(C_\ell^{\mathrm{kSZ}^2\delta_h } \right)^2\right\}\delta_{\ell\ell'} \,,
\end{equation}
where $\Delta\ell$ is the bin width, $C_\ell^{\Theta_f^2\Theta_f^2}$ is the auto-power spectrum of the filtered-squared CMB+noise+kSZ realization, $C_\ell^{\delta_h\delta_h}$ is the halo overdensity map auto-power spectrum, $N_\ell^{\delta_h\delta_h}$ is the halo shot noise power spectrum, and lastly $C_\ell^{\mathrm{kSZ^2\delta_h}}$ is the projected field kSZ--halo cross-power spectrum. Since the halo distribution is Poissonian, the shot noise term is
\begin{equation} \label{shot_noise}
N_\ell^{\delta_h\delta_h} = \frac{1}{\bar{n}_h},
\end{equation}
with $\bar{n}_h$ the number of halos per steradian.

To more precisely quantify the agreement between the theory and simulations, and to assess whether the analytic theory is sufficiently accurate for the interpretation of data (assuming the simulation curves are the ``truth'' that would be measured in data), we perform an amplitude fitting procedure, in which we fit the theoretical prediction to the map results using a $\chi^2$ minimization approach.  We scale the total halo model curve by an overall amplitude $A$, such that $A=1$ corresponds to perfect agreement between the analytic prediction and the simulations.  Using either approach to estimate the covariance (Monte Carlo or Gaussian approximation), we can define a Gaussian likelihood:
\begin{equation}
    \label{eq.chi2}
    -2 \ln \mathcal{L}(A) \equiv \chi^2(A) = \left(C_\ell^{\mathrm{kSZ}^2\delta_h, \mathrm{theory}} - C_\ell^{\mathrm{kSZ}^2\delta_h, \mathrm{sim.}} \right) \left( M_{\ell\ell'} \right)^{-1} \left(C_{\ell'}^{\mathrm{kSZ}^2\delta_h, \mathrm{theory}} - C_{\ell'}^{\mathrm{kSZ}^2\delta_h, \mathrm{sim.}} \right)
\end{equation}
We find that the Monte Carlo and Gaussian covariance approaches yield consistent results in the following analysis; therefore we adopt the Gaussian approximation for its computational efficiency.  Using the Gaussian likelihood, we find the best-fit amplitude $A$, as well as its error bar, that rescales the halo model curve to match the simulation data.  We emphasize that the purpose of this exercise is simply to determine the overall level of agreement between the theory and simulations, not to recalibrate the theory curves in any precise manner.  In addition, this exercise allows us to assess whether the analytic halo model prediction is sufficiently accurate for the interpretation of projected-field kSZ data from a given experiment: if $A$ deviates from unity at a statistically significant level, then an actual analysis using the analytic halo model for such an experiment would be biased.

The rescaled \verb|class_sz| curves are shown as red dashed lines in Figs.~\ref{ksz_so} and~\ref{ksz_planck}.  For SO, we obtain 
%$A = 0.838 \pm 0.0167$, $A = 0.789 \pm 0.0141$, and $A = 0.759 \pm 0.0209$ 
$A = 0.637 \pm 0.013$, $A = 0.662 \pm 0.011$, and $A = 0.662 \pm 0.017$
for the three redshift cuts (from lowest to highest redshifts, respectively).  These results indicate that \verb|class_sz| systematically overpredicts the simulation results at high significance, and that a projected-field SO kSZ analysis would yield biased results if the current halo model were used for the interpretation of the data.  For \emph{Planck}, we find 
%$A = 0.684 \pm 0.229$, $A = 0.641 \pm 0.312$, and $A = 0.722 \pm 0.730$.
$A = 0.564 \pm 0.164$, $A = 0.538 \pm 0.213$, and $A = 0.589\pm 0.462$. 
Due to the much larger error bars arising from the \emph{Planck} noise and beam as compared to SO, these values are statistically consistent with unity (except for some tension in the lowest redshift cut), implying reasonable agreement between the analytic model and the simulations for \emph{Planck}.  Thus, previous \emph{Planck}-based projected-field kSZ analyses~\cite{Hill2016,Kusiak_2021}, which relied on analytic models of the signal, are unlikely to have been biased at a statistically significant level due to limitations or approximations in the theoretical model, as also validated using different simulations in Ref.~\cite{Ferraro2016}.

\begin{figure}[t]
  \centering
  \includegraphics[width=.8\linewidth]{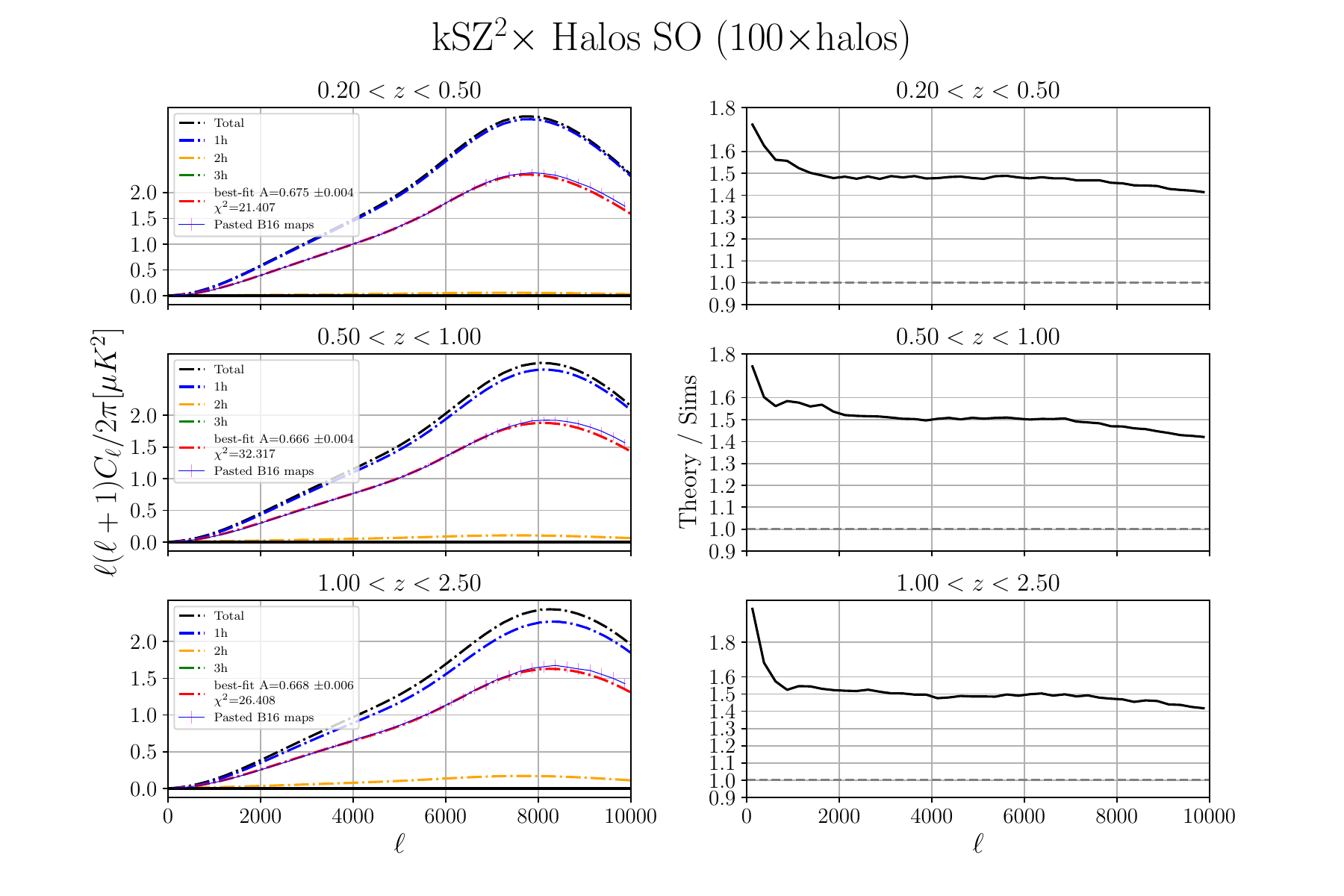}  
  \caption{Comparison between projected-field kSZ${}^2$--halo cross-correlation theory prediction and simulations for the SO setup, with error bars computed using shot noise 100 times lower than that in Fig.~\ref{ksz_so}, i.e., the number of halos is 100 times larger. The discrepancy between the theory and simulations is now more significant, with the uncertainties on $A$ reduced by a factor of 2-3 compared to those in Fig.~\ref{ksz_so}.}
\label{ksz_so_num}
\end{figure}

\begin{figure}[t]
  \centering
  \includegraphics[width=.8\linewidth]{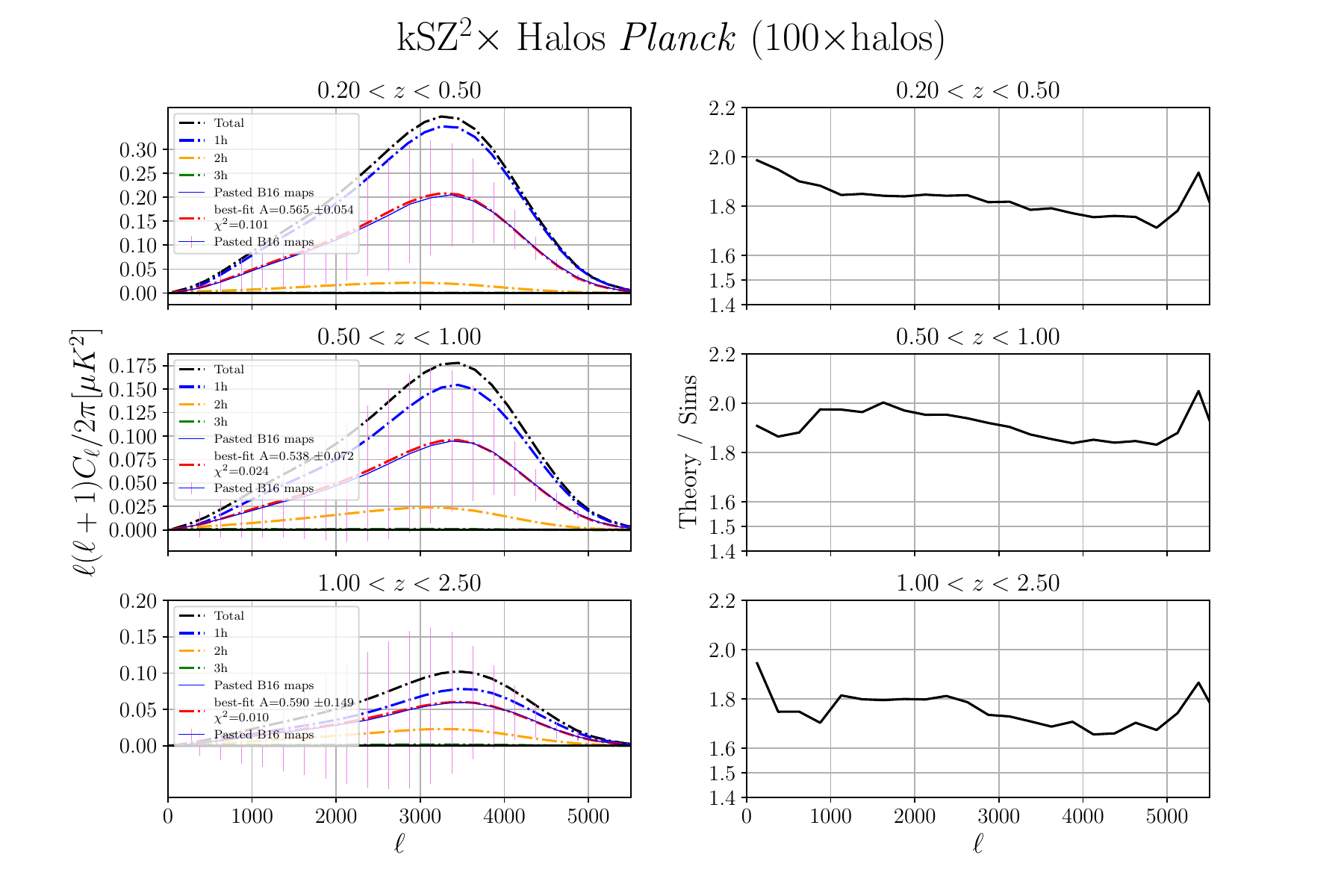}
  \caption{Comparison between projected-field kSZ${}^2$--halo cross-correlation theory calculations and simulations for the \emph{Planck} setup for three redshift cuts, with error bars computed using shot noise 100 times lower than that in Fig.~\ref{ksz_planck}, i.e., the number of halos is 100 times larger.  The \textit{Planck} simulation results are still in moderate agreement with our model despite the lower shot noise. } %, in particular on scales below $\ell\approx4000$. At $\ell \gtrsim 4000$ for the $0.2<z<0.5$ and $0.5<z<1$ cuts, our halo model is moderately discrepant with the simulation measurements.}}
  \label{ksz_planck_num}
\end{figure}

To more precisely assess the accuracy of our halo model calculation compared to the simulations, we recompute the error bars for realistic upcoming halo sample sizes.  Note that the number of halos in the three redshift cuts that we consider from Websky (see Table~\ref{halo_num_z}) are all much lower than the number of halos in already-existing galaxy catalogs (e.g., \emph{unWISE} contains hundreds of millions of galaxies~\cite{krolewski_2020}).  To recompute the error bars, we simply alter the shot noise by theoretically changing the number of halos in our sample.  By increasing the number of halos, we decrease the shot noise.  However, in order to do this, we need to approximately separate the signal and noise contributions to the total auto-power spectrum of each halo overdensity map.  We compute $C_{\ell, \mathrm{tot}}^{\delta_h\delta_h}$ by taking the auto-power spectrum of each halo overdensity map.  Approximating the shot noise as purely Poissonian as in Equation~\eqref{shot_noise}, we then subtract it from $C_{\ell, \mathrm{tot}}^{\delta_h\delta_h}$ and fit a simple power-law model to the large-scale signal-only halo power spectrum $C_{\ell}^{\delta_h\delta_h}$:
\begin{equation} \label{power_law}
C_{\ell}^{\delta_h\delta_h} \simeq A_1 (\ell/50)^{A_2} \,,
\end{equation}
where $A_1$ and $A_2$ are free parameters.  We perform the fit from $\ell = 2$ to $\ell = 100$ for all redshift cuts, where the signal term dominates over the shot noise (we also find that the shot noise can deviate from a purely Poissonian shape at higher $\ell$, such that subtracting it can occasionally lead to negative signal power on some angular scales, but on large scales this simple approach is accurate).  After determining $A_1$ and $A_2$ for each redshift cut, we then have a model of $C_{\ell}^{\delta_h\delta_h}$.  We can then add a new shot noise term to this model, with any desired number of theoretical halos in the sample, to obtain a new total halo auto-power spectrum to use in the covariance matrix in Equation~\eqref{gauss_cov}.  We use this approach to easily forecast measurements with larger halo sample sizes, as presented in Figs.~\ref{ksz_so_num} and \ref{ksz_planck_num}.  In particular, for simplicity we consider increasing the number of halos in each redshift cut by a factor of 100, yielding samples that are similar in size to \emph{unWISE} or other ongoing/upcoming surveys.

The \textit{Planck} results in Fig.~\ref{ksz_planck} show that the theoretical predictions fall reasonably in line with the pasted B16 simulation measurements.  Even after decreasing the halo shot noise in Fig.~\ref{ksz_planck_num} for halo sample sizes 100 times larger than those of Websky, we find that this conclusion still mostly holds.  %Some discrepancy is seen in the $0.2 < z < 0.5$ cut, particularly on small scales ($\ell \gtrsim 3500$).  The intermediate-redshift sample shows similar but slightly less discrepancy, while the $1 < z< 2.5$ sample agrees within the error bars on all angular scales.  The amplitude-fitting results also reflect this behavior, yielding 
%$A=0.680 \pm 0.084$, $A=0.638 \pm 0.163$, and $A=0.718 \pm 0.331$,
The amplitude fitting yields $A = 0.565 \pm 0.054$, $A = 0.538 \pm 0.072$, and $A = 0.590\pm 0.149$ for the three redshift cuts, 
respectively.  Thus, these results indicate that use of the analytic theory model for the interpretation of \textit{Planck} $\times$ \emph{unWISE} data is marginally acceptable, but that the use of the simulation-based theory would lead to a higher inferred amplitude in the analyses of Refs.~\cite{Hill2016,Kusiak_2021}.  Interestingly, this would increase tension between the inferred kSZ amplitudes in those analyses and theoretical expectations.  Given the large uncertainties from \emph{Planck}, we do not investigate this issue in detail here, but caution that further theoretical development is needed.

The SO results clearly reinforce this conclusion.  After decreasing the halo shot noise in Fig.~\ref{ksz_so_num} for halo sample sizes 100 times larger than those of Websky, the discrepancy between the theory and simulation curves in Fig.~\ref{ksz_so_num} is highly significant.  The best-fit rescaled amplitudes of the halo model power spectra are 
%$A=0.838 \pm 0.0056$, $A=0.790 \pm 0.0072$, and $A=0.759 \pm 0.0094$ 
$A = 0.675 \pm 0.004$, $A = 0.666 \pm 0.004$, and $A = 0.668\pm 0.006$ for $0.2<z<0.5$, $0.5<z<1.0$, and $1.0<z<2.5$, 
respectively.  Assuming that the primary cause of this discrepancy is the missing Wick contractions in the theory calculation, this indicates that the sub-leading terms contribute at the 20-30\% level, consistent with (but slightly larger than) the results of~\cite{Patki_2023}.  However, the near-scale-independence of the discrepancy that we find differs from the scale-dependent behavior seen in~\cite{Patki_2023}, which will require future investigation.  The overall takeaway is that improvements to the theoretical halo model for projected-field kSZ cross-correlations will be needed in order to obtain unbiased inferences from upcoming SO and galaxy survey data.

To further validate our results, we explore further variations of these calculations in Appendix~\ref{Appendix}, where we repeat the projected-field kSZ analysis using a different gas density profile.  %in Appendices~\ref{Appendix} and~\ref{appendix_b}.  In Appendix~\ref{Appendix}, we repeat the projected-field kSZ analysis using a different gas density profile; in Appendix~\ref{appendix_b}, we adopt simple Gaussian filters for $w(\ell)$, rather than using experiment-specific filters and beams.  
We briefly detail the results of Appendix~\ref{Appendix} here.  We choose new values for the GNFW gas parameters, as listed in Table~\ref{non_b16_gas}, and construct new kSZ maps with this profile.  We run the same analysis as presented above with the new kSZ maps, and compare the resulting cross-correlations with \verb|class_sz| predictions, accounting for the new parameter values. The results are shown in Appendix~\ref{Appendix} in Figs.~\ref{ksz_so_non_b16} and~\ref{ksz_planck_non_b16} for the SO and \emph{Planck} filters, respectively.  For SO, the ratios between the \verb|class_sz| predictions and simulations for the non-B16 gas profile similarly again show little scale dependence, but with overall values near $\approx 1.3$. For \emph{Planck}, the ratio in the two lowest redshift bins is roughly 1.5, but in the highest redshift bin it drops to 1.25.  This check further reinforces the conclusion that the differences that we find between the theory and simulations are real (rather than an artifact or a missing overall factor in the code), as they change significantly when the underlying physical model is changed. In addition, in Appendix~\ref{Appendix_b}, we present results for a third experimental filter, finding evidence of stronger scale dependence in the resulting theory-to-simulation ratio, thus suggesting again that the differences we see are not due to an artifact.

%however, the difference is significant, especially for the highest redshift bin ($1<z<2.5$), where the ratio is $\approx$ 0.80, i.e., the \verb|class_sz| prediction is lower than the cross-correlation measured from simulations. 

As a final calculation, we compute the signal-to-noise ratios (SNR) of the projected-field kSZ measurements as simulated by Websky for the  \emph{Planck} and SO experimental setups, given by
\begin{equation} \label{SNR}
    \mathrm{SNR} = \sqrt{ C_{\ell}^\mathrm{kSZ^2\delta_h} \left(M_{\ell\ell'}\right)^{-1}C_{\ell'}^\mathrm{kSZ^2\delta_h}} \,,
\end{equation}
where $M_{\ell\ell'}$ is the covariance in Equation~\eqref{gauss_cov}. 
We run the calculation for the Websky halo sample size (indicated by $N_h$) and 100 times the sample size ($100 \times N_h$) for each redshift and filter, and present the results in Table~\ref{SNR_results}.  Note that the $100N_h$ case is comparable to the size of existing galaxy catalogs like \emph{unWISE}, particularly for the $0.5 < z < 1.0$ redshift cut.  %Thus, the SNR forecast for \emph{Planck} and SO in this case should be realistic, and indeed for \emph{Planck} it is similar to the SNRs obtained with actual data~\cite{Hill2016,Kusiak_2021}.  Overall, 
The future SNR expected from SO with the projected-field estimator is very high (as found in~\cite{Ferraro2016,Patki_2023,Bolliet_2023}), thus motivating further theoretical development to improve the model used in interpreting the data.
% Ref.~\cite{Bolliet_2023} obtains similar SNR values for \textit{Planck} and SO using the \emph{unWISE} galaxy catalog \cite{Schlafly_2019}. \ok{i don't think we should compare to bolliet et al SNRs}

\begin{table}[t]
\begin{tabular}{ c|c|c|c|c| } 
Halo Sample Size & Experiment & $0.2<z<0.5$ & $0.5<z<1.0$ & $1.0<z<2.5$ \\ 
\hline
 %  $N_h$ & SO & $50.1$ & $54.0$ & $35.0$ \\  
 % & \textit{Planck} & $3.0$ & $2.0$ & $1.0$  \\  
 % \hline
 % $100\times N_h$ & SO & $159$ & $122$ & $87.1$ \\  
 % & \textit{Planck} & $8.9$ & $4.4$ & $2.5$ \\ 

   $N_h$ & SO & $53.6$ & $61.3$ & $38.6$ \\  
 & \textit{Planck} & $3.0$ & $2.0$ & $1.0$  \\  
 \hline
 $100\times N_h$ & SO & $163.2$ & $182.9$ & $119.8$ \\  
 & \textit{Planck} & $8.9$ & $4.4$ & $2.5$ \\ 
\end{tabular}
    \caption{Signal-to-noise ratios for the cross-correlation of the projected-field kSZ signal with halos, as measured in Websky, for two different experimental configurations corresponding to SO and \emph{Planck}, for the three different redshift bins considered in this work. We report SNR values for the baseline Websky halo samples (see Table~\ref{halo_num_z}), as well as for a hypothetical case where the halo sample size is increased by a factor of 100. The covariance used is an analytical Gaussian covariance given by Equation~\eqref{gauss_cov}, which includes contributions from the primary CMB, kSZ, and noise including the effects of residual foregrounds (note, however, that the \emph{Planck} SNR calculations here include only the primary CMB and kSZ contributions to the covariance).}
    \label{SNR_results}
\end{table}

%%%%%%%%%%%%%%%%%%%%%%%%%%%%%%%%%%%%%%%%%%%%%%%%%%%%%%%%%%%%%%%%%%%%

\section{Discussion}
\label{sec:discussion}

In this paper, we present a detailed comparison of the projected-field kSZ signal computed in the halo model using \verb|class_sz| with measurements from simulations.  Using three different subsets of halos from the Websky simulation suite as our LSS tracer, we find a difference of $\sim 50$-$80\%$ when compared with the analytic halo model calculation of the dominant term in the kSZ$^2$ $\times$ halos five-point function (see Equation~\eqref{bispec_approx}), as implemented in \verb|class_sz|, and described in~\cite{Bolliet_2023}.

As a validation of our computational setup, we first verify that simulated tSZ and $\tau$ maps yield halo cross-correlations that agree with our analytic halo model calculations, when implemented using the same profiles, halo redshift distributions, and parameters.  Although the tSZ and $\tau$ comparisons agree across all redshift cuts considered, we do not necessarily expect the same from the kSZ results, as our analytic model in this case only includes the dominant expected term.  As noted after Equation~\eqref{three-point_function}, 
% \begin{gather*} \label{5-point_func}
%  \langle \delta v \delta v \delta_h \rangle \sim \langle vv \rangle \langle \delta\delta\delta_h \rangle
%  + \langle \delta v \rangle \langle \delta v \delta_h \rangle + \langle \delta\delta \rangle \langle vv\delta_h \rangle \\ + 
%  \langle v\delta \rangle \langle v\delta\delta_h \rangle + 
%  \langle v\delta \rangle \langle \delta v\delta_h \rangle + 
%  \langle \delta v \rangle \langle v\delta\delta_h \rangle \\ + 
%  \langle v\delta_h \rangle \langle \delta\delta v \rangle + 
%  \langle \delta_hv \rangle \langle \delta\delta v \rangle + 
%  \langle \delta\delta_h \rangle \langle vv\delta \rangle + 
%  \langle \delta\delta_h \rangle \langle vv\delta \rangle,
% \end{gather*}
the $\langle vv\rangle\langle \delta_e\delta_e\delta_{h}\rangle$ term is most dominant in the five-point function, and this is the only term currently implemented in \verb|class_sz|.  In addition to the missing terms, our analytic calculation also makes standard halo model approximations, which may play a role in the comparison to simulations as well.

Our main results are presented in Sec.~\ref{subsec:kSZ}.  Interestingly, although there are terms missing in our analytic model, the shape of the angular power spectrum is very similar to that found from the simulation measurements.  However, the overall amplitude differs, with the theoretical prediction generally exceeding the simulation measurements. %(see Appendix~\ref{Appendix} for one case where the opposite behavior is found).  
For SO, the theory curve is generally  $\approx 50\%$ larger than the simulation power spectra, while for \emph{Planck}, the difference ranges from 70-100\%.  We verify that the near scale-independence of the theory-to-simulation ratios is not due to a missing factor in the projected-field calculation by directly comparing the result to that of a different profile in Appendix~\ref{Appendix}, where different ratio values are found and some scale dependence is seen for \emph{Planck}.

It is informative to compare our results with those of \cite{Patki_2023}, who developed an improved model for the projected-field kSZ signal including contributions from all terms in the Wick expansion, and also including a more rigorously derived version of the dominant $\langle vv \rangle \langle \delta_e \delta_e \delta_h \rangle$ term that differs from, and is less than or equal to, this term in the ``approximate model'' (the model used here).  We find that the amplitude of the numerical difference between our theory and simulations is at the expected level compared to the differences that \cite{Patki_2023} find between their improved model and the approximate model.  Although Ref.~\cite{Patki_2023} used the ``effective approach'' to model the projected-field kSZ signal as opposed to our halo model approach, their improved model shows a $\sim 10-20\%$ difference from the ``approximate model'', which only includes the $\langle vv\rangle\langle \delta_e\delta_e\delta_{h}\rangle$ contraction.  In our results, the \verb|class_sz| curves are analogous to their ``approximate model,'' in that we expect a similar difference from our theory curve to the simulations as they do with the improved model. They find that their improved model yields a higher signal for $\ell \lesssim 5500$ and a lower signal on smaller scales, compared to the approximate model (for SO).  In our analysis, however, the simulation results show a lower cross-correlation than that of our analytic model on all scales. In addition, their ratio plots are clearly angular-scale-dependent, showing the most discrepancy around $\ell\approx1500$ for \textit{Planck} and around $\ell\approx3000$ and $\ell \approx 8000$ for SO.  The near-scale-independence of our results in the right panels of Figs.~\ref{ksz_so} and~\ref{ksz_planck} remains a curiosity, but may be explained by properly combining the ``improved model'' of Ref.~\cite{Patki_2023} with the halo model developed in Ref.~\cite{Bolliet_2023}, which we defer to future work.  Indeed, the high-$\ell$ behavior that we find for SO in comparing the simulations to the theory is similar to that seen in~\cite{Patki_2023} in comparing their improved model to the approximate model.

This work could be improved further by considering a more realistic setup that includes foregrounds in the simulated sky maps, such as tSZ, CIB, or CMB lensing (which contributes to the signal through the so-called ``lensing term'' that arises since squaring a filtered temperature map reconstructs the lensing potential in a non-optimal way~\cite{Hill2016,Ferraro2016}). Different mitigation strategies as discussed in, e.g., \cite{Kusiak_2021} could then be tested to isolate the blackbody kSZ component and test the fidelity of the measured signal.  In particular, it would be valuable to explore component separation techniques as seen in \cite{surrao2024constrainingcosmologicalparametersneedlet} and \cite{mccarthy2024signalpreservingcmbcomponentseparation} with foreground-contaminated maps.  In this work, we have only focused on biases arising from incomplete or inaccurate modeling of the signal itself, but in practice foreground biases are of even larger concern in projected-field kSZ measurements~\cite{Hill2016,Kusiak_2021}.

In addition, recent work~\cite{Patki2024} introduced a new bispectrum estimator to measure the kSZ effect, analogous to the projected-field estimator but probing the full three-point function.  In short, instead of squaring a temperature map, Ref.~\cite{Patki2024} computes the full three-point correlation function $\langle T T \delta_h \rangle$.  By avoiding the convolution introduced by the squaring operation, this method enables cleaner scale separation than the original projected-field kSZ estimator. The forecast SNR for a combination of SO and \emph{unWISE} yields $\approx106 \sigma$ (similar in magnitude to our findings in Table~\ref{SNR_results}). Testing this novel estimator on simulations, as done here for the simpler projected-field cross-power spectrum estimator, would also be an interesting direction for future work.

%%%%%%%%%%%%%%%%%%%%%%%%%%%%%%%%%%%%%%%%%%%%%%%%%%%%%%%%%%%%%%

\begin{acknowledgments}
We thank Raagini Patki and Nick Battaglia for useful conversations and comments on the manuscript, and Boris Bolliet for help with \verb|class_sz|. We acknowledge computing resources from Columbia University's Shared Research Computing Facility project, which is supported by NIH Research Facility Improvement Grant 1G20RR030893-01, and associated funds from the New York State Empire State Development, Division of Science Technology and Innovation (NYSTAR) Contract C090171, both awarded April 15, 2010.  MJR, AK, and JCH acknowledge support from NSF grant AST-2108536.  MJR acknowledges additional support from the Columbia Bridge to the Ph.D.~Program.  JCH acknowledges additional support from the Sloan Foundation and the Simons Foundation.  Several software packages were used in the analysis described in this paper, including \verb|HEALPix/healpy| \cite{Healpix, Healpy}, \verb|numpy| \cite{numpy}, \verb|scipy| \cite{scipy}, \verb|matplotlib| \cite{matplotlib}, and \verb|astropy| \cite{astropy1, astropy2, astropy3}.
\end{acknowledgments}

%%%%%%%%%%%%%%%%%%%%%%%%%%%%%%%%%%%%%%%%%%%%%%%%%%%%%%%%%%%%%%

\appendix

\section{Results for Alternative Gas Density Profiles}
\label{Appendix}
\begin{figure}[t]
  \centering
  \includegraphics[width=.8\linewidth]{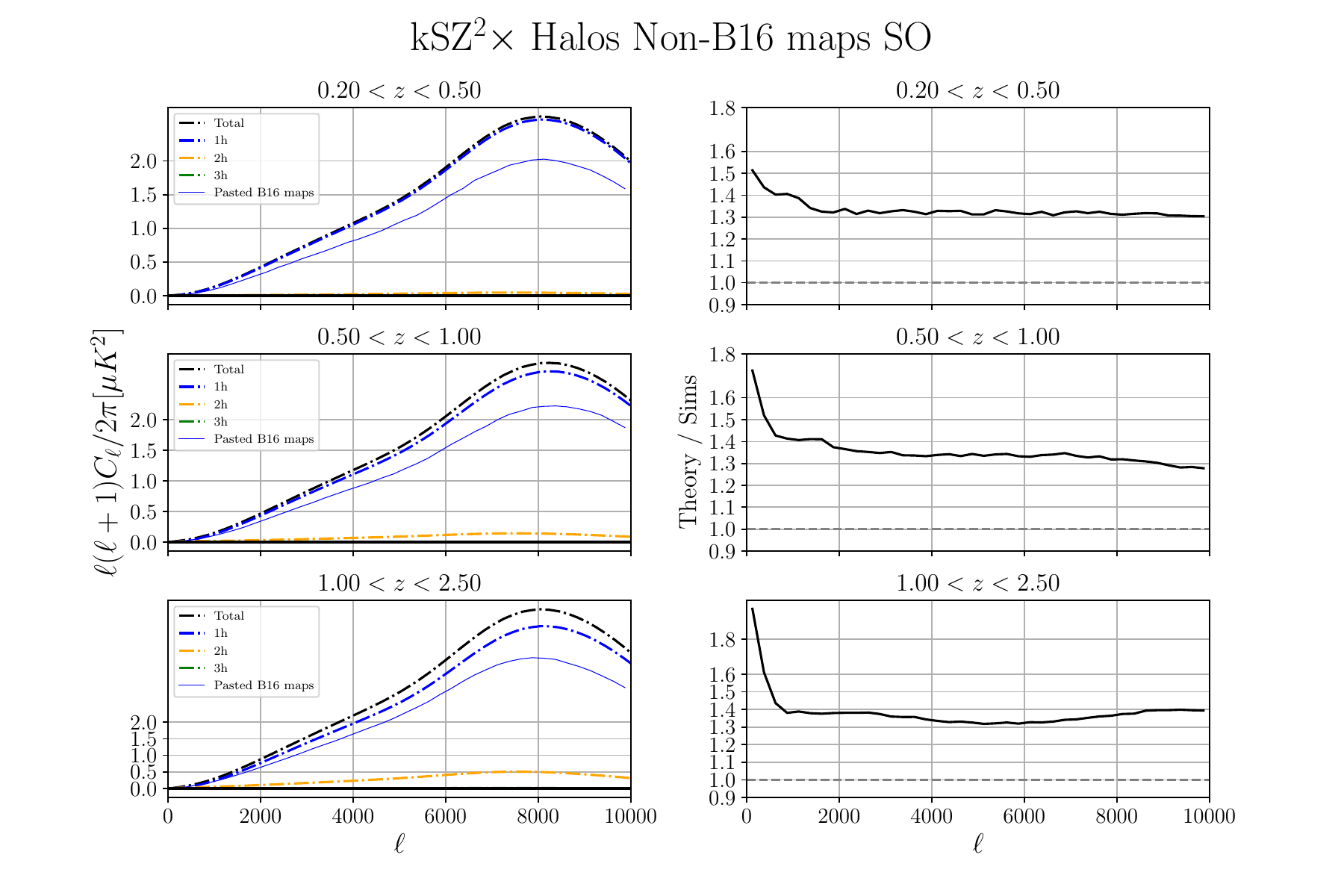}
  \caption{Projected-field kSZ$^2\times$ halos with non-B16 electron density profile. We compare the theory and simulation results for SO (left) and show the theory to simulation ratio (right).  These results are similar to our main results in Fig.~\ref{ksz_so}: the theory-to-simulation ratio is still nearly scale-independent, but is slightly lower than that found in Fig.~\ref{ksz_so}.}
  \label{ksz_so_non_b16}
\end{figure}

\begin{figure}[t]
  \centering
  \includegraphics[width=.8\linewidth]{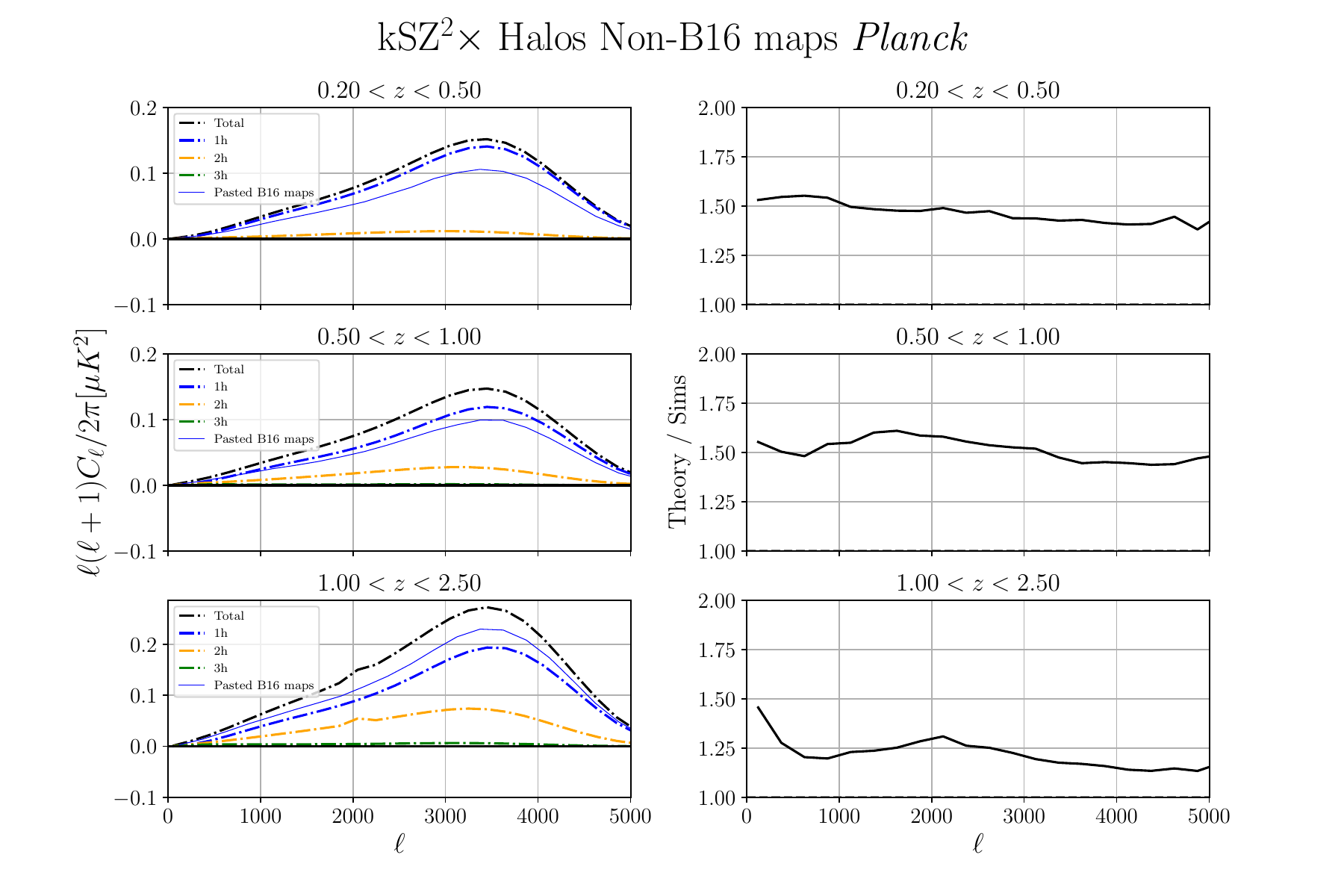}
  \caption{Projected-field kSZ$^2\times$ halos with non-B16 electron density profile.  We compare the theory and simulation results for \textit{Planck} (left) and show the theory to simulation ratio (right). Here, we see clear distinctions from our results in Fig.~\ref{ksz_planck}. The $1<z<2.5$ cut is most notable, with the theory matching the simulations within a factor of 1.25, as compared to 1.5 in the lower redshift bins.} %this is the only instance in which \texttt{class\_sz} under-predicts the simulation signal, resulting in a theory to map ratio of $C_{\ell}^{\mathrm{theory}}/C_{\ell}^{\mathrm{maps}} \approx 0.8$ over most of the multipole range.}}
  \label{ksz_planck_non_b16}
\end{figure}

In addition to using the B16 gas density profile, we also construct simulated kSZ maps with a different set of parameters to test our conclusions. It is peculiar that for the fiducial results, when we compute the ratios between the theory and pasted maps, across filters and redshifts, the ratios have very little scale dependence and are generally close to 1.5 (although they do change somewhat with redshift).  As a sanity check, we run the same projected-field calculation in Figs.~\ref{ksz_so_non_b16} and  \ref{ksz_planck_non_b16} with the parameters listed in Table~\ref{non_b16_gas}.

\begin{table}[t]
\begin{tabular}{ c | c c c }
  & $A_0$ & $A_M$ & $A_z$ \\ 
  \hline
 $C$ & $5\times10^3$ & $0.4$ & $-0.8$ \\  
 $\alpha$ & $0.88$ & $-0.03$ & $0.19$ \\   
 $\beta$ & $4.83$ & $0.25$ & $-0.5$
\end{tabular}
    \caption{GNFW profile parameters used for non-B16 electron density profile.}
\label{non_b16_gas}
\end{table}

With this alternative gas profile, we find that the ratios between theory predictions and simulations change compared to those found in Sec.~\ref{subsec:kSZ}, but still show very little scale dependence. The most notable change compared to the fiducial GNFW results is for the \textit{Planck} $1 < z < 2.5$ bin. 
% This is the only case we find where the \verb|class_sz| theory prediction falls below the simulation measurement, with a ratio $\approx 0.8$. The $0.5 < z < 1$ bin for \textit{Planck} exhibits some mild $\ell$ dependence, steadily increasing to past $1.25$. 
For SO, the results remain broadly similar to those obtained with the fiducial GNFW profile in Fig.~\ref{ksz_so}, but cluster around  $\approx 1.3$-$1.4$ rather than $1.5$.  However, the lack of scale dependence persists.  Overall, these results provide strong evidence that our theory calculation is not simply missing an overall factor compared to the simulations, and indicate that the differences we find are genuine limitations of the current halo model calculation.

\section{Results for Alternative Filter}
\label{Appendix_b}

\begin{figure}[t]
    \centering
    \includegraphics[width=0.8\linewidth]{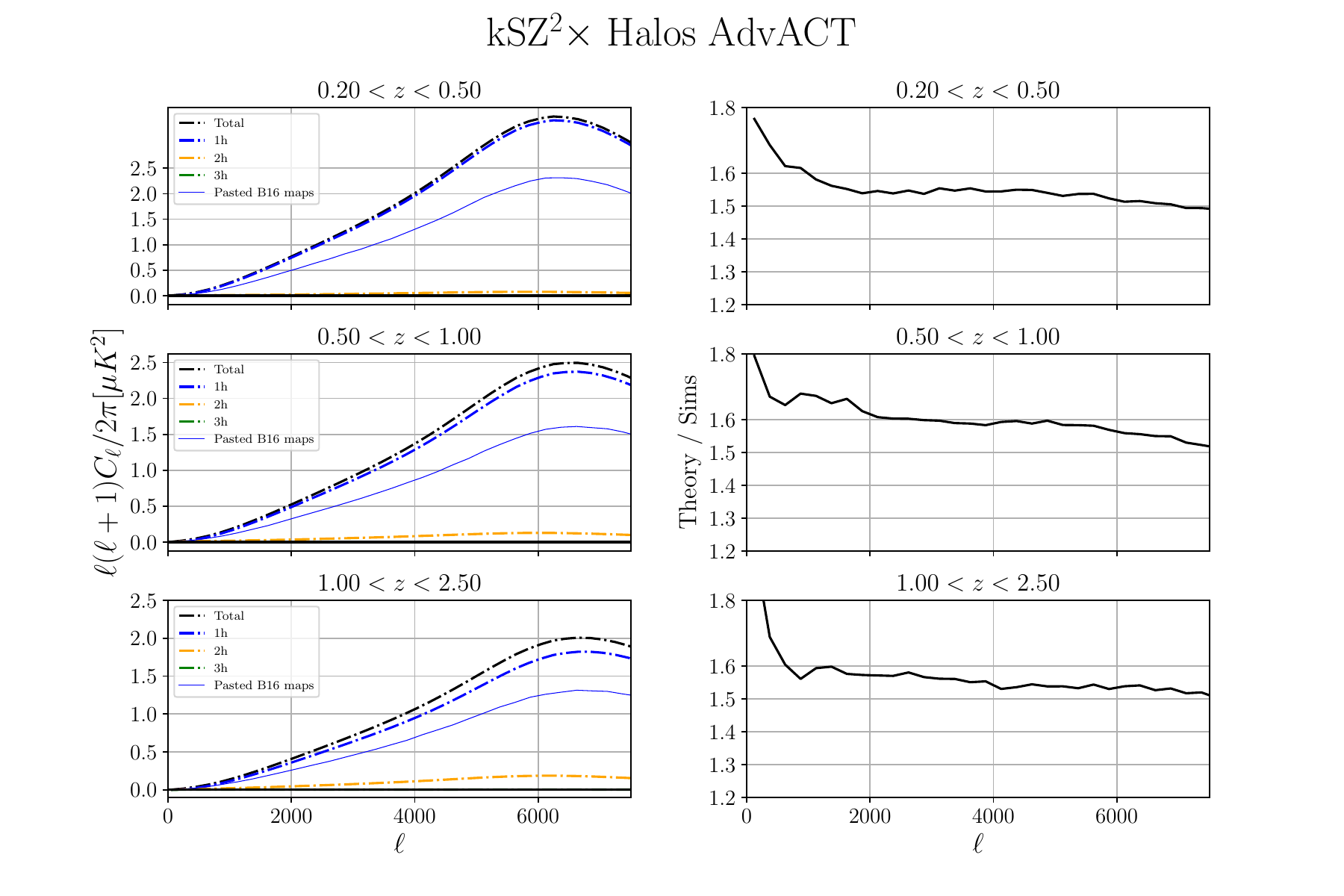}
    \caption{Comparison between projected-field kSZ${}^2$--halo cross-correlation theory calculations and simulations for the AdvACT filter (see Fig.~\ref{filters_plot}). In the left panels, the dash-dotted black lines denote the total halo-model prediction, comprised of the one-halo (dashed-dotted blue), two-halo (dashed-dotted orange), and three-halo (dashed-dotted green) terms. The thin blue lines with error bars show the results measured from simulations, with errors computed via the Gaussian covariance approximation.  In the right panels, we show the corresponding theory-to-simulation ratios, which are close to $\approx 1.55$, though they show more scale-dependence than the SO results. }
    \label{fig:ksz2xhalos_act}
\end{figure}

In this appendix, we show results for another filter constructed for the Advanced ACT (AdvACT) experimental setting, as presented in Ref.~\cite{Bolliet_2023}. We show the results in Fig.~\ref{fig:ksz2xhalos_act}.

For this setup, we see a similar trend of the \verb|class_sz| prediction being larger than the simulation measurements, with ratios $\approx 1.5-1.6$ for all three redshift bins, though showing more scale dependence than the main SO results presented in Fig.~\ref{ksz_so}.

\bibliography{main}

\end{document}